\pdfoutput=1
\RequirePackage{ifpdf}
\ifpdf 
\documentclass[pdftex]{sigma}
\else
\documentclass{sigma}
\fi

\usepackage[geometry]{ifsym}

\numberwithin{equation}{section}

\def\ii{{\rm i}}
\def\dd{{\rm d}}
\def\'#1{{\accent19\ifx #1i \i\else #1\fi}}
\def\onehalf{\textstyle\frac12}
\def\tsty#1#2{{\textstyle\frac{#1}{#2}}}
\def\ssr#1{{\scriptscriptstyle\rm #1}}

\def\Poi#1#2{\{#1,#2\}_\ssr{P}}
\def\Lie#1{\hbox{\sf #1}}
\def\cuadro{{\!\lower3pt\hbox{\Square}}}
\def\cuadrito{{\lower2pt\hbox{\scriptsize\Square}}}
\def\cutito{{\lower2pt\hbox{\tiny\Square}}}
\def\rombo{{\!\lower3pt\hbox{\Diamondshape}}}
\def\rombito{{\lower2pt\hbox{\scriptsize\Diamondshape}}}
\def\ketsub#1#2#3{\,|#1\rangle_{\!\ssr{#2}}^{#3}}
\def\brasub#1#2#3{\phantom{{|}}_{\ssr{#2}}^{{}\!\! #3}\!\langle {#1}|}
\def\braketsub#1#2#3#4#5#6{\phantom{{|}e}_{\ssr{#2}}^{{}\!\! #3}\!\langle
		{#1}|{#4}\rangle_{\ssr{#5}}^{#6}}

\def\be{\begin{equation}}
\def\ee{\end{equation}}
\def\bea{\begin{eqnarray}}
\def\eea{\end{eqnarray}}

\def\hboxa#1#2#3{\framebox{\vbox to#1{\vfil\hbox to#2{\hfil
		$\displaystyle #3$\hfil}\vfil}}\hskip-0pt\null}
\def\vboxa#1#2#3{\null\vskip-10.8pt\vbox to#1{\vss\hbox to#2{\hfil
		$\displaystyle #3$\hfil}\vss}\vskip-11.8pt\null}
\def\hcaja#1{\hboxa{0.7cm}{0.7cm}{#1}}
\def\vcaja#1{\vboxa{0.7cm}{3cm}{#1}}
\def\Hcaja#1{\hboxa{0.7cm}{1.8cm}{#1}}
\def\Vcaja#1{\vboxa{0.7cm}{7cm}{#1}}
\def\patternthree#1#2#3{%
\vbox{\vcaja{\hfill\hcaja{#1}\hcaja{#2}}\vcaja{\hfill\hcaja{#3}}}}

\def\patternfour#1#2#3#4#5#6{%
\vbox{\vcaja{\hfill\hcaja{#1}\hcaja{#2}\hcaja{#3}}
\vcaja{\hfill\hcaja{#4}\hcaja{#5}}
\vcaja{\hfill\hcaja{#6}}}}
\def\Patternfour#1#2#3#4#5#6{%
\vbox{\Vcaja{\hfill\Hcaja{#1}\Hcaja{#2}\Hcaja{#3}}
\Vcaja{\hfill\Hcaja{#4}\Hcaja{#5}}
\Vcaja{\hfill\Hcaja{#6}}}}

\begin{document}

\allowdisplaybreaks

\renewcommand{\thefootnote}{$\star$}

\renewcommand{\PaperNumber}{053}

\FirstPageHeading

\ShortArticleName{The Fourier U(2) Group and Separation of Discrete Variables}

\ArticleName{The Fourier U(2) Group and Separation\\ of Discrete Variables\footnote{This paper is a
contribution to the Special Issue ``Symmetry, Separation, Super-integrability and Special Functions~(S$^4$)''. The
full collection is available at
\href{http://www.emis.de/journals/SIGMA/S4.html}{http://www.emis.de/journals/SIGMA/S4.html}}}

\Author{Kurt Bernardo WOLF~$^\dag$ and Luis Edgar VICENT~$^{\ddag}$}

\AuthorNameForHeading{K.B.~Wolf and L.E.~Vincent}

\Address{$^\dag$~Instituto de Ciencias
 F\'{\i}sicas, Universidad Nacional Aut\'onoma de M\'exico,\\
\hphantom{$^\dag$}~Av.\ Universidad s/n, Cuernavaca, Mor.\ 62210, M\'exico}
\EmailD{\href{mailto:bwolf@fis.unam.mx}{bwolf@fis.unam.mx}}
\URLaddressD{\url{http://www.fis.unam.mx/~bwolf/}}

\Address{$^{\ddag}$~Deceased}

\ArticleDates{Received February 19, 2011, in f\/inal form May 26, 2011;  Published online June 01, 2011}

\Abstract{The linear canonical transformations of geometric optics on
two-dimensional screens form the group \Lie{Sp($4,\Re$)}, whose
maximal compact subgroup is the Fourier group $\Lie{U($2$)}_\ssr{F}$;
this includes isotropic and anisotropic Fourier transforms, screen
rotations and gyrations in the phase space of ray positions and
optical momenta.  Deforming classical optics into a Hamiltonian
system whose positions and momenta range over a f\/inite set of values,
leads us to the f\/inite oscillator model, which is ruled by the Lie
algebra \Lie{so($4$)}.  Two distinct subalgebra chains are used to
model arrays of $N^2$ points placed along Cartesian or polar (radius
and angle) coordinates, thus realizing one case of separation in
two discrete coordinates. The $N^2$-vectors in this space are
digital (pixellated) images on either of these two grids, related by
a unitary transformation. Here we examine the unitary action of the
analogue Fourier group on such images, whose rotations are
particularly visible.}


\Classification{20F28; 22E46; 33E30; 42B99; 78A05; 94A15}

\section{Introduction}   \label{sec:one}

The real symplectic group \Lie{Sp($4,\Re$)} of linear canonical
transformations is widely used in paraxial geometric optics
\cite[Part 3]{GeomOpt} on two-dimensional screens, and in classical
mechanics with quadratic potentials. Its maximal compact subgroup is
the {\it Fourier\/} group $\Lie{U($2$)}_\ssr{F}
=\Lie{U($1$)}_\ssr{F}\otimes \Lie{SU($2$)}_\ssr{F}$
\cite{Simon-KBW1,Simon-KBW2}. These are classical Hamiltonian systems
with two coordinates of position $(q_x,q_y)\in\Re^2$, and two
coordinates of momentum $(p_x,p_y)\in\Re^2$, which form the phase
space $\Re^4$, and which are subject to linear canonical
transformations that preserve the volume element.  The central
subgroup $\Lie{U($1$)}_\ssr{F}$ contains the fractional isotropic
Fourier transforms, which rotate the planes $(q_x,p_x)$ and
$(q_y,p_y)$ by a common angle.  The complement group
$\Lie{SU($2$)}_\ssr{F}$ contains anisotropic Fourier transforms that
rotate the planes $(q_x,p_x)$ and $(q_y,p_y)$ by opposite angles; it
also contains joint rotations of the $(q_x,q_y)$ and $(p_x,p_y)$
planes; and thirdly, gyrations, which `cross-rotate' the planes
$(q_x,p_y)$ and $(q_y,p_x)$.  The rest of the \Lie{Sp($4,\Re$)}
transformations shear or squeeze phase space, as free f\/lights and
lenses, or harmonic oscillator potential jolts.

This classical system can be quantized {\it\`a la Schr\"odinger\/}
into paraxial wave or quantum models. Indeed, the group of canonical
integral transforms was investigated early by Moshinsky and Quesne in
quantum mechanics \cite{Mosh-Quesne1,Mosh-Quesne2}, \cite[Part
4]{IntTfms} and by Collins in optics \cite{Collins}, and yielded the
integral transform kernel that unitarily represents the (two-fold
cover of the) group \Lie{Sp($4,\Re$)} on the Hilbert space ${\cal
L}^2(\Re^2)$. Yet it is the {\it discrete\/} version of this system
which is of interest for its technological applications to sensing
wavef\/ields with {\sc ccd} arrays. One line of research addresses the
discretization of the integral kernel to a matrix and its computation
using the fast FFT algorithm
\cite{Ding,Hennely-Sheridan,Ozaktas-group,Healy-Sheridan}.
This has the downside that the kernel matrices are in general not
unitary, and thus no longer represent the group~-- non-compact groups
can have only inf\/inite-dimensional unitary representations
\cite{Gilmore}.

We have developed a `{\it finite quantization\/}' process to pass
between classical quadratic Hamiltonian systems, to systems whose
position and momentum coordinates are discrete and f\/inite
\cite{AW97,AVW99,APVW-I,Echaya,Sohag}. The set of values a discrete
and f\/inite coordinate can have is the spectrum of the generator of a
compact subalgebra of \Lie{u($2$)}, which is a deformation of the
oscillator Lie algebra \Lie{osc}. When position space is
two-dimensional, we use $\Lie{so($4$)}=\Lie{su($2$)}_x \oplus
\Lie{su($2$)}_y$ \cite{Sohag,APVW-II,VW08}. The purpose of this
article is to elucidate the action of the Fourier group
$\Lie{U($2$)}_\ssr{F}$ on the $N^2$-dimensional representation spaces
of the algebra \Lie{so($4$)} that we use to realize pixellated
images.

\looseness=1
Previously we have analyzed and
synthesized pixellated images on sensor arrays that follow Cartesian
or polar coordinates \cite{APVW-II,VW08,rot-LEV};
here we investigate the action of the Fourier
group on f\/inite pixellated images arranged along these two coordinate
systems, understood as one example of separation of coordinates in
discrete, f\/inite spaces. In order to present a reasonably
self-contained review of our approach to two-dimensional f\/inite
systems, we must repeat developments that have appeared previously,
which will be recognized by the citation of the relevant references.
As research experience indicates however, each restatement of
previous results yields a streamlined and better structured text, where
the notation is unif\/ied and directed toward the economic statement of
the solution to the problem at hand. In the present case, it is the
action of the four-parameter Fourier group on Cartesian and
polar-pixellated images.

The classical Fourier subgroup of paraxial optics on two-dimensional
screens~\cite{Simon-KBW2} is described in Section~\ref{sec:two}. To
f\/initely quantize this, we review the one-dimensional case, where
\Lie{u($2$)} is used to model the f\/inite harmonic oscillator
\cite{AW97,Echaya,Sohag} in Section~\ref{sec:three}.  The ascent to
two-dimensional f\/inite systems, Cartesian and polar, occupies the
longer Section \ref{sec:four}. There, under the {\ae}gis of
\Lie{so($4$)} we describe its f\/inite position space separated in
Cartesian~\cite{APVW-I} and in polar~\cite{APVW-II} coordinates,
together with the two-parameter subgroups of the Fourier group that
are within \Lie{so($4$)} for each coordinate system. The new
developments begin in Section~\ref{sec:five}, where we import the
continuous group of rotations on images pixellated along Cartesian
coordinates from their natural action on the polar pixellation. This
serves to complete the action of the $\Lie{U($2$)}_\ssr{F}$ Fourier
group on Cartesian screens in Section \ref{sec:six}. To translate the
action of this group onto polar screens, we recall in Section~\ref{sec:seven} the unitary transformation between Cartesian and
polar-pixellated screens~\cite{VW08}, thus f\/inding the representation
of the Fourier group on the circular grid as well. Finally, in
Section~\ref{sec:eight} we of\/fer some comments on the wider context
in which we hope to place the separation of discrete coordinates in
two dimensions.

\section{The classical Fourier group}   \label{sec:two}

The classical oscillator is
characterized by the Lie algebra \Lie{osc} of Poisson brackets
between the oscillator Hamiltonian function $h(q,p):=
\frac12(p^2+q^2)$ and the phase space coordinates of position~$q$ and
momentum~$p$,
\begin{gather}
	\Poi hq = -p, \qquad \Poi hp = q, \qquad \Poi qp := 1,
				\label{class-Ham}
\end{gather}
where 1 Poisson-commutes with all. The f\/irst two brackets are
the geometric and dynamic Hamilton equations, while the last one
actually def\/ines the algebra of the system to be \Lie{osc}.

\begin{figure}[t]
\centering 
\includegraphics[width=2.2in]{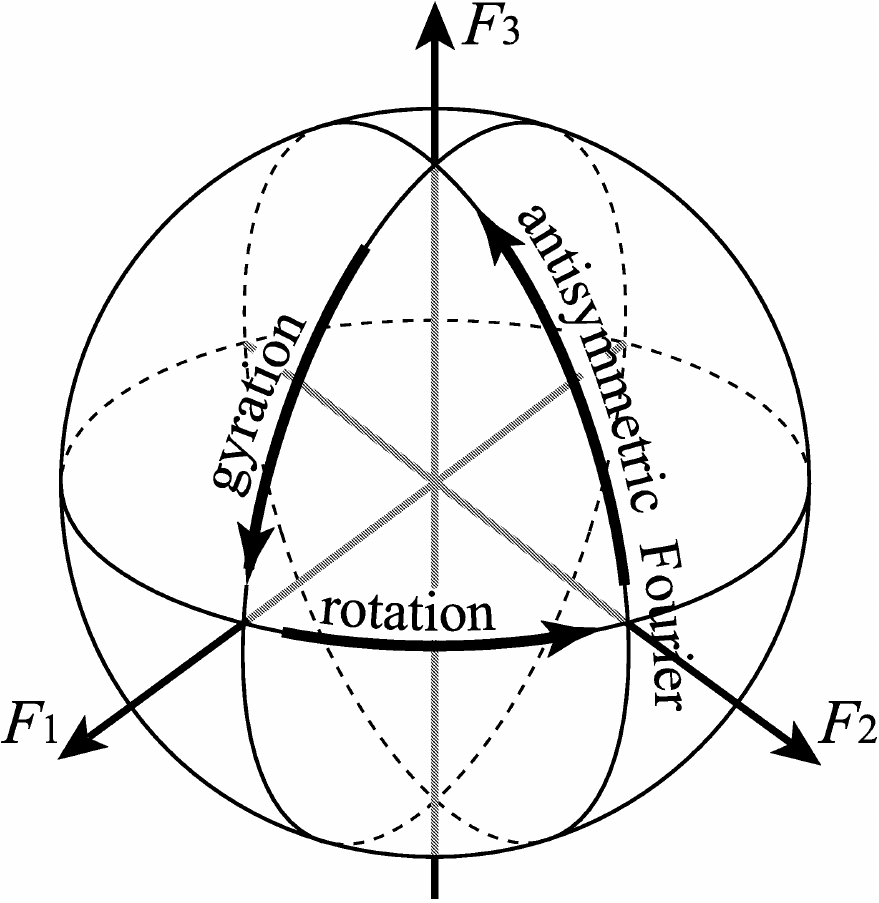}
\caption{Action of the Fourier group on the sphere (from~\cite{Madrid}).}
\label{fig:UFourier}
\end{figure}

In two dimensions $(q_x,q_y)\in\Re^2$ and with the corresponding
momenta $(p_x,p_y)\in\Re^2$, one can build ten independent quadratic
functions
\begin{gather}  q_iq_{i'},\qquad q_ip_{i'},\qquad p_ip_{i'},\qquad
         i,{i'}\in\{x,y\},   \label{ten-gens}
\end{gather}
that will close into the real symplectic Lie algebra
\Lie{sp($4,\Re$)}. Four linear combinations of them generate
transformations that include the fractional Fourier transform (FT),
which are of interest in optical image processing,
\begin{gather}
	\hbox{isotropic FT \ } F_0 :=
		\tsty14\left(p^2_x+p^2_y+q^2_x+q^2_y\right)=:\onehalf(h_x+h_y),
			\label{symmetricFT}\\
	\hbox{anisotropic FT \ } F_1 :=
		\tsty14\left(p^2_x-p^2_y+q^2_x-q^2_y\right)=:\onehalf(h_x-h_y),
			\label{skewsymmetricFT}\\
	\hbox{gyration \ } F_2 :=  \tsty12(p_xp_y+q_xq_y),
			\label{gyration}\\
	\hbox{rotation \ } F_3 :=  \tsty12(q_xp_y-q_yp_x).
			\label{rotation}
\end{gather}
Their Poisson brackets close within the set,
\begin{gather*}
	\Poi{F_1}{F_2}=F_3, \qquad \hbox{and cyclically}, \\ 
	\Poi{F_0}{F_i}  = 0 , \qquad i=1,2,3,   
\end{gather*}
and characterize the {\it Fourier\/} Lie algebra
$\Lie{u($2$)}_\ssr{F}= \Lie{u($1$)}_\ssr{F}\oplus
\Lie{su($2$)}_\ssr{F}$, with $F_0$  central. This is the
maximal compact subalgebra of \eqref{ten-gens},
$\Lie{u($2$)}_\ssr{F} \subset \Lie{sp($4,\Re$)}$  \cite{Simon-KBW2}.
In Fig.~\ref{fig:UFourier} we show the transformations generated by
\eqref{skewsymmetricFT}--\eqref{rotation} on the linear space of
the \Lie{su($2$)} algebra, which leaves invariant the spheres
$F^2_1+F^2_2+F^2_3 = F^2_0$.

The Lie algebra \Lie{u($2$)} and group \Lie{U($2$)} will appear in
several guises, so it is important to distinguish their `physical'
meaning. The classical Fourier group $\Lie{U($2$)}_\ssr{F}$ of linear
optics will be matched by a `Fourier--Kravchuk' group
$\Lie{U($2$)}_\ssr{K}$ acting on the position space of f\/inite
`sensor' arrays in the f\/inite oscillator model. In one dimension,
this model is based on the algebra \Lie{u($2$)}, whose generators are
position, momentum, and energy; while in two, the algebra is
$\Lie{su($2$)}_x\oplus \Lie{su($2$)}_y=\Lie{so($4$)}$. In all, the
well-known properties of $\Lie{su($2$)}=\Lie{so($3$)}$ will of use~\cite{Bied-Louck}.

\section{Finite quantization in one dimension}
							         \label{sec:three}

The usual Schr\"odinger quantization $(q\mapsto\bar Q=q\cdot$,
$p\mapsto\bar P=-\ii\eta \dd/\dd q)$ of the phase space coordinates
leads to the paraxial model of wave optics (for the reduced
wavelength $\eta=\lambda/2\pi$) or to oscillator quantum mechanics
(for $\eta=\hbar$). In one dimension, \eqref{ten-gens} yields the three
generators of the (double cover of the) group \Lie{Sp($2,\Re$)} of
canonical integral transforms \cite{Mosh-Quesne1} acting on functions~-- continuous inf\/inite signals~-- in the Hilbert space ${\cal
L}^2(\Re)$. In two dimensions, the Schr\"odinger quantization of
\eqref{ten-gens} yields the generators of \Lie{Sp($4,\Re$)}, and its
Fourier subgroup is generated by the quadratic Schr\"odinger
operators $\bar F_0$, $\bar F_1$, $\bar F_2$, $\bar F_3$
corresponding to \eqref{symmetricFT}--\eqref{rotation}, that we indicate
with bars, and are up-to-second order dif\/ferential operators acting
on the Hilbert space.  {\it Finite quantization\/} on the other hand,
asigns self-adjoint $N\times N$ {\it matrices\/} $q\mapsto Q$,
$p\mapsto P$ to the coordinates of phase space.

Since the oscillator algebra is noncompact, it cannot have a faithful
representation by f\/inite self-adjoint matrices \cite{Gilmore}; we must thus
deform the four-generator \Lie{osc} into a compact algebra, of which
there is a single choice: $\Lie{u($2$)}$.  We should keep the
geometry and dynamics of the harmonic oscillator contained in the two
Hamilton equations in \eqref{class-Ham}, with commutators
$\ii[\,\cdot\,,\,\cdot\,]$ in place of the Poisson brackets
$\Poi{\cdot}{\cdot}$, and in place of the oscillator Hamiltonian $h$,
a matrix $K$ added with some constant times the unit matrix. We
call $K$ the {\it pseudo\/}-Hamiltonian. The
third~-- and fundamental~-- commutator $[Q,P]$ will then give back $K$.
In \cite{AW97} we proposed the assignments of self-adjoint $N\times
N$ matrices given by the well-known irreducible representations of
the Lie algebra \Lie{su($2$)} of spin $j$, thus of dimension
$N=2j+1$, where the matrix representing position is chosen diagonal.
Using the generic notation $\{J_1,J_2,J_3\}$ for generators of
\Lie{su($2$)}, the matrices are
\begin{gather}
	\hbox{position:} \quad q\mapsto Q=J_1, \nonumber\\
	\qquad Q_{m,m'} = m\delta_{m,m'},\qquad m,m'\in
				\{-j,-j+1,\ldots, j\},
								\label{matQ}\\
	\hbox{momentum:} \quad p\mapsto P=J_2, \nonumber\\
\qquad	P_{m,m'} = -\ii\onehalf\sqrt{{( j-m)( j+m+1)}}
									\delta_{m+1,m'}  +\ii\onehalf\sqrt{{( j+m)( j-m+1)}}
									\delta_{m-1,m'}, \label{matP}\\
	\hbox{pseudo-energy:} \quad K=J_3,\qquad\hbox{energy:} \quad
							h\mapsto H:=K+(j+\onehalf)1, \nonumber\\
\qquad	K_{m,m'} = \onehalf\sqrt{{( j-m)( j+m+1)}}
									\delta_{m+1,m'}  +\onehalf\sqrt{{( j+m)( j-m+1)}}
									\delta_{m-1,m'}. \label{matL}
\end{gather}
The commutation relations of these matrices are
\begin{gather}
[K,Q]=-\ii P,\qquad [K,P]=\ii Q, \qquad [Q,P]=-\ii K, \nonumber\\
	\hbox{i.e.}, \qquad [J_1,J_2]=-\ii J_3, \qquad
				\hbox{and cyclically}.  	\label{commRel}
\end{gather}
The central generator 1 of \Lie{osc} is corresponded with the
unit $N\times N$ matrix that generates a~\Lie{U($1$)} central
group of multiplication by overall phases. This completes the
algebra $\Lie{u($2$)}$ realized by matrices acting on the linear space
${\cal C}^N$ of complex $N$-vectors.

The spectra $\Sigma$ of the matrices $Q$ of position $\{q\}$, $P$ of
momentum $\{\varpi\}$, and $K$ of pseudo-energy $\{\kappa\}$ are
thus discrete and f\/inite,
\begin{gather*}
	\Sigma(Q)=\Sigma(P)=\Sigma(K)=\{-j,-j+1,\ldots,j\},
\end{gather*}
with $j$ integer or half-integer. We def\/ine the {\it mode number\/}
as $n:=\kappa+j\in\{0,1,\ldots,2j\}$, and {\it energy\/} is
$n+\onehalf$. The Lie group generated by $\{1;Q,P,K\}$ is the
\Lie{U($2$)} group of linear unitary transformations of ${\cal C}^N$.
Finally, the sum of squares of these matrices on ${\cal C}^N$ is the
Casimir invariant
\begin{gather*}
	C:=Q^2+P^2+K^2=j(j+1) 1.    
\end{gather*}
In Dirac notation, ${\cal C}^N$ is a spin-$j$ \Lie{u($2$)}
representation, where the eigenstates of position $q$ and of mode
$n=\kappa+j$ satisfy
\begin{gather}
	Q\ketsub{q}{1}{}=q\ketsub{q}{1}{},\qquad
	K\ketsub{n}{3}{}=(n-j)\ketsub{n}{3}{},\qquad
		q|_{-j}^j,\qquad n|_0^{2j}.
					\label{eigenf-pos-psenergy}
\end{gather}
The position eigenvectors $\ketsub{q}{1}{}$ form a Kronecker basis
for ${\cal C}^N$ where the components of $\ketsub{n}{3}{}$
will be provided, from \eqref{matL}, by a three-term relation, i.e., a dif\/ference equation that is satisf\/ied by the Wigner {\it
little-d\/} functions \cite{Bied-Louck} for $\kappa=n-j$. (The reader
may be disconcerted for having $J_1$ diagonal, whereas almost
universally $J_3$ is declared {\it the\/} diagonal matrix; formulas
obtained by permuting $1\mapsto2\mapsto3\mapsto1$ are unchanged.)
The overlaps between the two bases in \eqref{eigenf-pos-psenergy} are
the f\/inite oscillator wavefunctions,
\begin{gather}
	\Psi_n(q) := \!\!\!\!\braketsub{q}1{}{n}3{} = d^j_{n-j,q}\big(\onehalf\pi\big)
				=\Psi_{q+j}(n-j)
				\label{Psi-braket-d}	\\
\phantom{\Psi_n(q)}{} \,		 =  \frac{(-1)^n}{2^j}
		\sqrt{\left({2j\atop n}\right)\left({2j\atop q+j}\right)}
		K_n\big(q+j;\onehalf,2j\big).  \label{Kravch-func}
\end{gather}
This overlap is expressed here \cite{Atak-Suslov} in terms of the
square root of a binomial coef\/f\/icient in $q$, which is a discrete
version of the Gaussian, and a symmetric Kravchuk polynomial of
degree $n$, $K_n(q)\equiv K_n(q;\onehalf,N-1) = K_q(n-j)$
\cite{Krawtchouk}.  We have called the $\Psi_n(q)$'s  {\it Kravchuk
functions\/} \cite{AW97}; they form a real, orthonormal and complete
basis for ${\cal C}^N$. The Lie exponential of the self-adjoint
matrix $K$ generates the  unitary \Lie{U($1$)} group of {\it
fractional Fourier--Kravchuk\/}  transforms, whose realization we
shall indicate by $\Lie{U($1$)}_\ssr{K}$.

For future use we give the general expression of the Wigner
little-$d$ functions for angles $\beta\in[0,\pi]$ as a
trigonometric polynomial \cite{Gilmore,Bied-Louck}
\begin{gather}
	d^j_{m',m}(\beta)  =
		 \sqrt{(j+m')! (j-m')! (j+m)! (j-m)!}\nonumber \\
\phantom{d^j_{m',m}(\beta)  =}{}\times \sum_k
		\frac{(-1)^{m'-m+k} (\cos\onehalf\beta)^{2j+m-m'-2k}
		(\sin\onehalf\beta)^{m'-m+2k} }{
		k! (j+m-k)! (m'-m+k)! (j-m'-k)!} \label{func-d} \\
\phantom{d^j_{m',m}(\beta)}{}		 =  d^j_{m',m}(-\beta) = (-1)^{m-m'}d_{j-m,j-m'}(\beta),
					\label{reflect}
\end{gather}
where due to the denominator factorials, the summation extends over
the integer range of $\max (0,m-m')\le k\le
\min (j-m',j+m)$; for $m-m'>0$ the ref\/lection formulas~\eqref{reflect} apply.

\section{Two-dimensional systems}              \label{sec:four}

The two-dimensional classical oscillator algebra $\Lie{osc}_2$ is
sometimes taken to consist of the Poisson brackets between $1$,
$q_x$, $q_y$, $p_x$, $p_y$, and $h_x$, $h_y$; and sometimes an angular momentum
$m=q_1p_2-q_2p_1$ is included, Poisson-commuting with the total Hamiltonian
$h=h_x+h_y$. The f\/inite quantization of two-dimensional systems
deforms $\Lie{osc}_2$ to the Lie algebra $\Lie{su($2$)}\oplus
\Lie{su($2$)}=\Lie{so($4$)}$ in both cases. (The subalgebra
of 1 need not concern us here.) To work comfortably
with the six generators of $\Lie{so($4$)}$, we consider the customary
realization of $J_{i,i'}=-J_{i',i}$ by self-adjoint operators that
generate rotations in the $(i,i')$ planes, which obey the
commutation relations
\begin{gather*}
	[J_{i,i'},J_{k,k'}] = \ii(\delta_{i',k}J_{i,k'}
			+ \delta_{i,k'}J_{i',k}
			+ \delta_{k,i}J_{k',i'}
			+ \delta_{k',i'}J_{k,i}),
\end{gather*}
that we  summarize with the following pattern:

\vspace{2mm}

\begin{equation}
\vcenter{\patternfour{J_{1,2} }{J_{1,3} }{J_{1,4} }{J_{2,3} }{J_{2,4}
		}{J_{3,4}}}\ .	 \label{sopat4}
\end{equation}
A generator $J_{i,i'}$ has {\it non\/}-zero commutator with all those
in its row $i$ and its column $i'$ (ref\/lected across the $i=i'$
line); all its other commutators are zero. The asignments of
position, momentum and pseudo-energy with the
$\Lie{su($2$)}=\Lie{so($3$)}$ matrices \eqref{matQ}--\eqref{matL} is

{}\bigskip

\vbox{\baselineskip=14pt\[ {}\hskip-1.5cm{}
\vcenter{\patternthree{K}{{-}P}{Q}}
			{\qquad\hbox{\LARGE$=$}}
\vcenter{\patternthree{J_{1,2}}{J_{1,3}}{J_{2,3}}}
			{\qquad\hbox{\LARGE$=$}}
\vcenter{\patternthree{J_{3}}{{-}J_{2}}{J_{1}}}\ .
\]}

\subsection{The Cartesian coordinate system}

\noindent Passing from one to two dimensions can be achieved
{\it prima facie\/} by building the direct sum algebra
$\Lie{su($2$)}_x\oplus \Lie{su($2$)}_y$, which is accidentally equal
to \Lie{so($4$)}. This isomorphism is shown in patterns by

{}\bigskip
\vbox{\baselineskip=14pt\be{}\hskip-1.5cm{}
\vcenter{\patternthree{K^{\cutito}_x}{{}\!{-}P^{\cutito}_x}{Q^{\cutito}_x}}
			{\quad\hbox{\LARGE$\oplus$}\hskip-20pt}
\vcenter{\patternthree{K^{\cutito}_y}{{}\!{-}P^{\cutito}_y}{Q^{\cutito}_y}}
			{\quad\hbox{\LARGE$=$}\hskip-30pt}
\vcenter{\Patternfour{K^{\cutito}_x+K^{\cutito}_y
				}{-P^{\cutito}_x-P^{\cutito}_y
				}{Q^{\cutito}_x-Q^{\cutito}_y
   }{Q^{\cutito}_x+Q^{\cutito}_y}{P^{\cutito}_x-P^{\cutito}_y
					}{K^{\cutito}_x-K^{\cutito}_y}}\ ,\quad{}
							 \label{direct-sum} \ee}

\noindent where the square super-index of all generators, $X^{\cutito}_i$,
indicates the identif\/ication between the \Lie{so($4$)} generators
$J_{i,i'}$ with the Cartesian coordinates and observables.

\begin{figure}[t]
\centering 
\includegraphics[width=70mm]{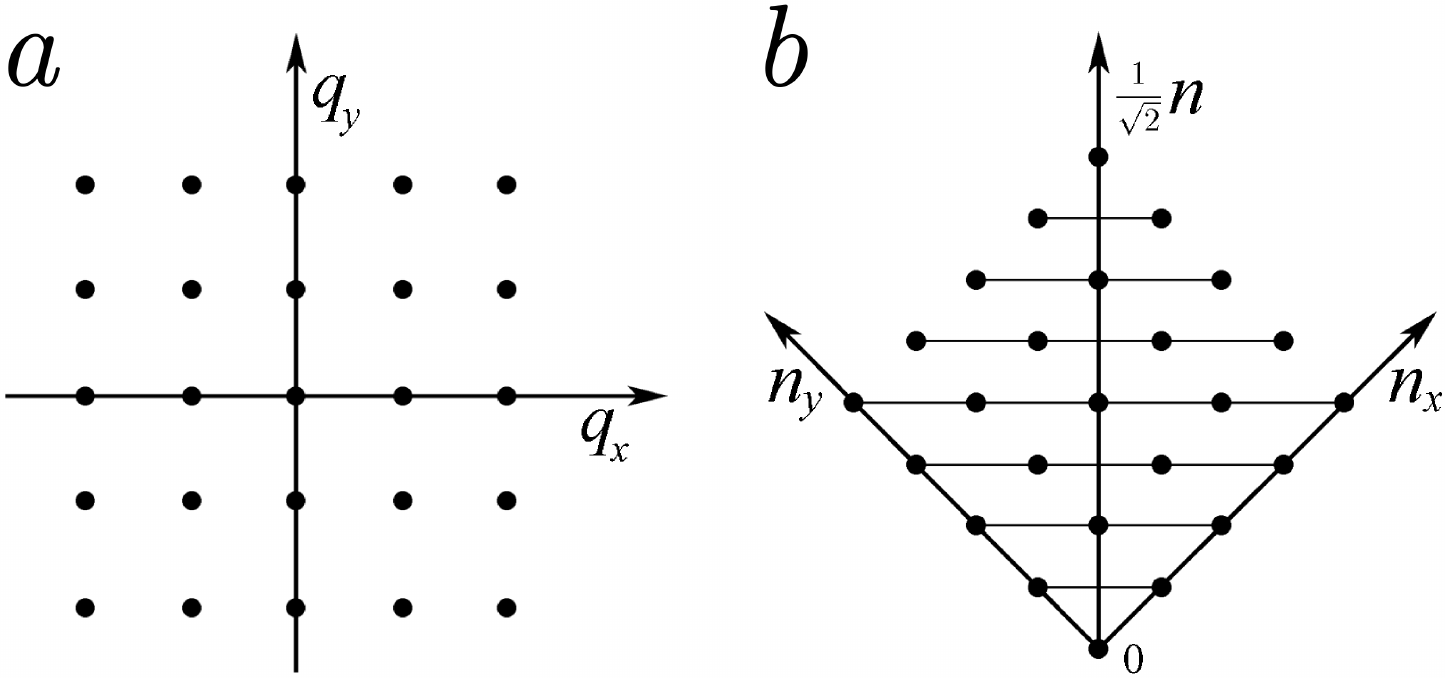}
\caption{Cartesian eigenvalues following the asignments of position
and energy generators in \eqref{direct-sum} within the algebra
\Lie{so($4$)}.  ({\it a}): Position eigenvalues $(q_x,q_y)$ in
\eqref{base-qxqy}.  ({\it b}): sum and dif\/ference of mode eigenvalues
$(n_x,n_y)$ in \eqref{base-kxky}.}
\label{fig:cuadrado-y-rombo}
\end{figure}

\begin{figure}[t]
\centering 
\includegraphics[width=2.0in]{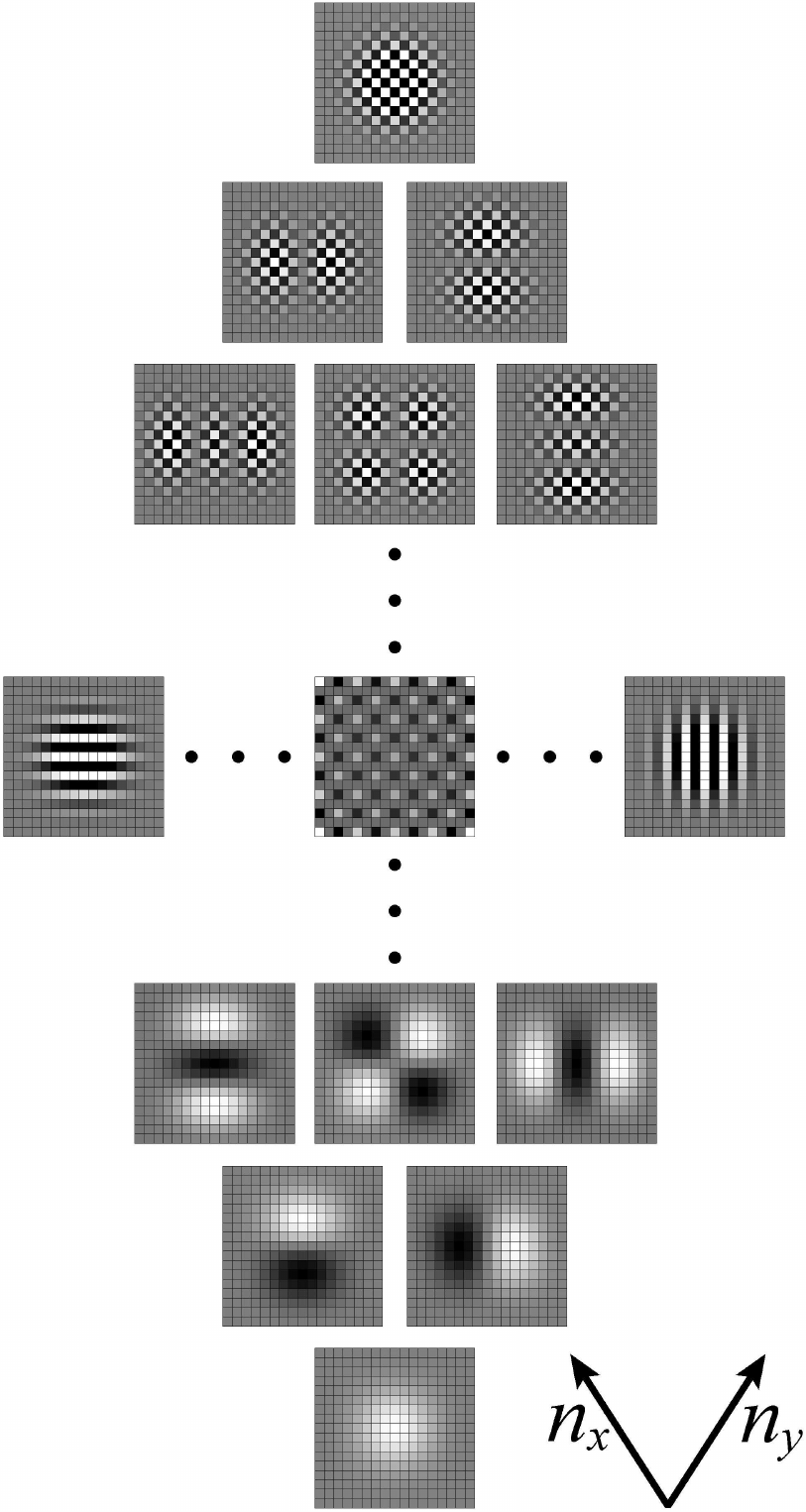}
\caption{Basis of $x$- and $y$-mode eigenstates on the Cartesian
array, $\Psi^\cutito_{n_x,n_y}(q_x,q_y)$ in \eqref{Twodimpsi}.}
\label{fig:base-Cartesian}
\end{figure}

Since the $x$-generators commute with the $y$-generators in
\eqref{direct-sum}, using \eqref{eigenf-pos-psenergy} a
Cartesian basis of positions and also a basis of modes
can be simply def\/ined as direct products
$\displaystyle\ketsub{q_x,q_y}{1\ }{\cutito} :=
\ketsub{q_x}{1x}{}\!\ketsub{q_y}{1y}{}$ and
$\displaystyle\ketsub{n_x,n_y}{3\ }{\cutito} :=
\ketsub{n_x}{3x}{}\!\ketsub{n_y}{3y}{}$,
\begin{alignat}{3}
&	Q^{\cutito}_x\ketsub{q_x,q_y}{1}{\cutito}
		=q_x\ketsub{q_x,q_y}{1}{\cutito}, \qquad &&
	Q^{\cutito}_y\ketsub{q_x,q_y}{1}{\cutito}=q_y\ketsub{q_x,q_y}{1}{\cutito},&
				\label{base-qxqy}\\
&	K^{\cutito}_x\ketsub{n_x,n_y}{3}{\cutito}
			=(n_x-j)\ketsub{n_x,n_y}{3}{\cutito},\qquad &&
	K^{\cutito}_y\ketsub{n_x,n_y}{3}{\cutito}
			=(n_y-j)\ketsub{n_x,n_y}{3}{\cutito},&
							\label{base-kxky}
\end{alignat}
where the pseudo-energy eigenvalues are $n_i-j=\kappa_i|_{-j}^j$.
The two-dimensional f\/inite oscillator wavefunctions in Cartesian
coordinates are thus given as a product of two $\Psi_n(q)$'s
from~\eqref{Psi-braket-d}, \eqref{Kravch-func},
\begin{gather}
	\Psi^\cutito_{n_x,n_y}(q_x,q_y)
	= \braketsub{q_x,q_y}{1}{\cutito}{n_x,n_y}{3}{\cutito}
						=\Psi_{n_x}(q_x) \Psi_{n_y}(q_y),
									\label{Twodimpsi}
\end{gather}								
on positions $q_x,q_y|_{-j}^j$ and of mode numbers $n_x,n_y|_0^{2j}$.
The positions can be accomodated in the square pattern of Fig.~\ref{fig:cuadrado-y-rombo}{\it a}, and the modes in the rhombus
pattern of Fig.~\ref{fig:cuadrado-y-rombo}{\it b}. The Cartesian
eigenstates of the f\/inite oscillator are shown in Fig.~\ref{fig:base-Cartesian}.

Since $K^{\cutito}_x$ and $K^{\cutito}_y$ generate independent
rotations in the $(Q_x,P_x)$ and in the $(Q_y,P_y)$ planes
respectively, their sum $K:=K^{\cutito}_x+K^{\cutito}_y=J_{1,2}$
transforms phase space isotropically. Thus we identify $K\in
\Lie{so($4$)}$ as the generator of a group $\Lie{U($1$)}_\ssr{K}$ of
isotropic fractional Fourier--Kravchuk transforms in
${\cal C}^N\times{\cal C}^N = {\cal C}^{N^2}$, and corresponding
with the operator $2\bar F_0\in\Lie{U($2$)}_\ssr{F}$ in~\eqref{symmetricFT}.  That sum commutes with $A:=K^{\cutito}_x-
K^{\cutito}_y= J_{3,4}\in\Lie{so($4$)}$, which generates
skew-symmetric Fourier rotations by opposite angles in the $x$- and
$y$- phase planes; so $A$ corresponds with $2\bar F_1 \in
\Lie{u($2$)}_\ssr{F}$ in~\eqref{skewsymmetricFT}.  However, a
counterpart of the $2\bar F_3$ generator of rotations in the
$(q_x,q_y)$ and $(p_x,p_y)$ planes {\it cannot\/} be found within
\Lie{so($4$)}, and neither can the gyrations in \eqref{gyration}. These
will be {\it imported\/} in the next section.

\subsection{The polar coordinate system}

The six generators of the Lie algebra \Lie{so($4$)} can be identif\/ied
with positions and modes following an asignment dif\/ferent from the
Cartesian direct sum of the previous section.  We indicate the new
generators with a circle super-index, $X^\circ_i$; as in the
classical case, a generator of rotations~$M$ between the $x$- and
$y$-axes should satisfy the commutation relations
\begin{alignat}{3}
& [M,Q^\circ_x]=\ii Q^\circ_y, \qquad && [M,Q^\circ_y]=-\ii Q^\circ_x,&
				\label{M-qxqy}\\
& [M,P^\circ_x]=\ii P^\circ_y, \qquad && [M,P^\circ_y]=-\ii P^\circ_x, &
				\label{M-pxpy}
\end{alignat}
while the isotropic Fourier generator $K=J_{1,2}$ should rotate
between position and momentum operators,
\begin{alignat}{3}
& [K,Q^\circ_x]=\ii P^\circ_x, \qquad &&
		[K,P^\circ_x]=-\ii Q^\circ_x, & \label{F-qxpx}\\
& [K,Q^\circ_y]=\ii P^\circ_y, \qquad &&
		[K,P^\circ_y]=-\ii Q^\circ_y, \qquad \hbox{and} &\label{F-qypy}\\
& [K,M]=0.  &&& \label{FMeq0}
\end{alignat}
The commutator~\eqref{FMeq0} asserts that $K$ and $M$ can be used to def\/ine a basis for
${\cal C}^{N^2}$ with the quantum numbers of mode and angular momentum,
that will be given below. Expressed in the pattern~\eqref{sopat4}, a new
\Lie{so($4$)} generator assignment that fulf\/ills these requirements is

{}\bigskip
\be
\vcenter{\patternfour{K}{\!{-}P^\circ_x}{\!{-}P^\circ_y
		}{Q^\circ_x}{Q^\circ_y}{M}} \;.	 \label{circpat4}
\ee
This assignment satisf\/ies the conditions \eqref{M-qxqy}--\eqref{FMeq0}, but
implies the further commutators
\begin{gather} [Q^\circ_x,P^\circ_x]   = \ii K = [Q^\circ_y,P^\circ_y], 	\label{Fnon-st-so4}\\
 [Q^\circ_x,Q^\circ_y]  = \ii M =			[P^\circ_x,P^\circ_y], 	\label{Mnon-st-so4}\\
  [Q^\circ_x,P^\circ_y]  = 0 =
			[Q^\circ_y,P^\circ_x]. \nonumber	
\end{gather}
Of these, \eqref{Fnon-st-so4} echoes the \Lie{su($2$)} nonstandard
commutator in~\eqref{commRel}, while the commutator~\eqref{Mnon-st-so4} is
also nonstandard, and indicates that $Q^\circ_x$ and $Q^\circ_y$
cannot be simultaneously diagonalized.

First, we f\/ind operators with quantum numbers corresponding to radius
and angle; we use the subalgebra chain

{}\bigskip
\vbox{\baselineskip=14pt\be\hskip1.5cm{}
\vcenter{\patternfour{K}{\!{-}P^\circ_x}{\!{-}P^\circ_y
		}{Q^\circ_x}{Q^\circ_y}{M}}	
			{\quad\hbox{\LARGE$\supset$}\hskip-20pt}
\vcenter{\patternthree{Q^\circ_x}{Q^\circ_y}{M}}	
			{\quad\hbox{\LARGE$\supset$}}
\vcenter{\hcaja{M}}\hss
			 \label{subalg-chain}\ee}

\begin{figure}[t]
\centering 
\includegraphics[width=70mm]{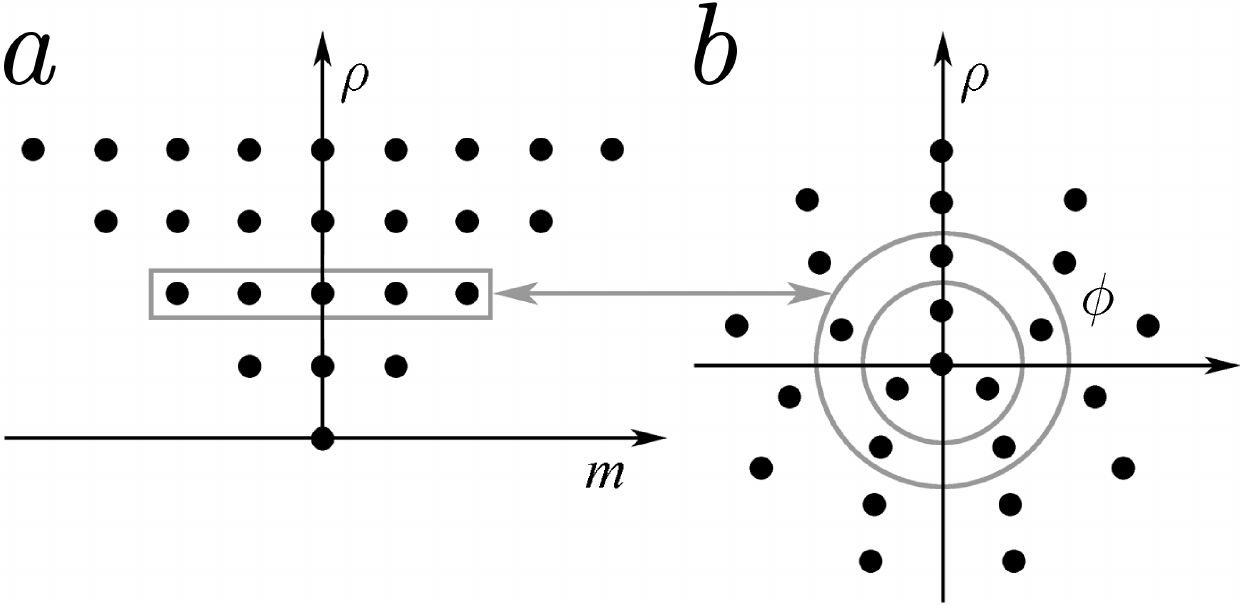}
\caption{Polar eigenvalues following the asignments of radius
and angle in \eqref{rho-phi-basis} within the algebra reduction
$\Lie{so($4$)}\supset\Lie{so($3$)}\supset\Lie{so($2$)}$ in~\eqref{subalg-chain}.
 {\it Left\/}:  Eigenvalues $(\rho,m)$ in \eqref{RM-eigenbasis}.
{\it Right\/}: Positions of radius~$\rho$ and angles~$\phi_k$
according to \eqref{rho-phi-basis} aligned by $\psi_\rho=0$.}
\label{fig:cono-y-circulo}
\end{figure}

\noindent When both \Lie{su($2$)}'s in \eqref{direct-sum} have the same
Casimir eigenvalue $j(j+1)$, the principal Casimir invariant is
$\sum\limits_{i<i'} J_{i,i'}^2=2j(j+1)1$.  In \Lie{so($4$)} there is a
second invariant,  $\sum\limits_{(i<i')\neq(k<k')}J_{i,i'}J_{k,k'}$,
which is identically zero in these `square' cases \cite{Bied-Louck}.
Let us now consider the  Casimir operator of the subalgebra
$\Lie{so($3$)}\subset\Lie{so($4$)}$  in \eqref{subalg-chain}, which is
\begin{gather}
	R(R+1):=(Q^\circ_x)^2 + (Q^\circ_y)^2 + M^2.
					\label{erre-cuad}
\end{gather}
The Gel'fand--Tsetlin branching rules \cite{Gelfand-Tsetlin} determine that
${\cal C}^{N^2}$ then decomposes into subspaces~${\cal C}^\rho$ that are
irreducible under  this \Lie{so($3$)}, where~\eqref{erre-cuad} exhibits the
eigenvalues \cite{Bied-Louck}
\[
	\rho(\rho+1),\qquad \rho\in\{0,1,\ldots,j-1\}.
\]
Although $R$ is {\it not\/} an element of the algebra \Lie{so($4$)},
we shall identify it as the {\it radius\/} operator.  The f\/inal link in the
reduction \eqref{subalg-chain} is $M$, whose eigenvalues in the
$\rho$-representation of \Lie{so($3$)} are $m\in\{-\rho,
-\rho+1,\ldots,\rho\}$. The total number of distinct eigenvalue
pairs $(\rho,m)$ is $\sum\limits_{\rho=0}^{2j} (2\rho+1)=(2j+1)^2=N^2$,
the same as for the $N\times N$ square grid, and shown in Fig.~\ref{fig:cono-y-circulo}{\it a}. Thus we def\/ine the radius-angular
momentum ({\sc ra}) eigenvectors
\begin{gather}
	R\ketsub{\rho,m}{RA}{\circ}
		=	\rho\ketsub{\rho,m}{RA}{\circ}, \qquad
	M\ketsub{\rho,m}{RA}{\circ}
		=	m\ketsub{\rho,m}{RA}{\circ},\qquad
			\rho|_0^{2j},\qquad m|_{-\rho}^\rho,
				\label{RM-eigenbasis}
\end{gather}
and we adopt $\rho$ as the radial position coordinate.

In the last step we use the discrete Fourier transform
matrix, ${\bf F}_\rho$ in each $(2\rho + 1)$-dimensional subspace, to
pass between  angular momenta and angles, and thus build the Kronecker
basis of states $\ketsub{\rho,\phi}{\odot}{}$ localized at a def\/inite
radius $\rho$ and the $2\rho+1$ equidistant angles $\phi_k$ as
\begin{gather}
	\ketsub{\rho,\phi_k}{\odot}{}:=\frac1{\surd 2\rho+1}
			\sum_{m=-\rho}^{\rho} \exp(-\ii m \phi_k)\,
				\ketsub{\rho,m}{RA}{\circ},
					\label{rho-phi-basis}\\
	\ketsub{\rho,m}{RA}{\circ} = \frac1{\surd 2\rho+1}
			 \sum_{k=-\rho}^{\rho} \exp(+\ii m \phi_k)\,
			\ketsub{\rho,\phi_k}{\odot}{}
\nonumber\\
	\hbox{for} \quad \phi_k:=2\pi k/(2\rho+1) + \psi_\rho,\qquad
				-\rho\le k\le \rho,  \nonumber
\end{gather}
where the $\psi_\rho$'s are f\/ixed but arbitrary phases. In Fig.~\ref{fig:cono-y-circulo}{\it b\/} we show the resulting arrangement
of $N^2$ points $(\rho,\phi_k)$ thus def\/ined. Note that the Fourier
transformation in \eqref{rho-phi-basis} is linear and unitary but is
{\it not\/} an element of the group \Lie{SO($4$)}: it has been {\it
imported\/} to act on each of the $2\rho+1$-dimensional \Lie{so($3$)}
irreducible representation subspaces $\rho\in\{0,1,\ldots,2j\}$.

Having the position Kronecker eigenstate basis
$\ketsub{\rho,\phi_k}{\odot}{}$ for polar coordinates, we now def\/ine
the basis of mode and angular momentum {\sc ma},  eigenbasis
of the commuting operators $K=K^{\cutito}_x+K^{\cutito}_y$ with total
mode number $n|_0^{4j}$ (pseudo-energy $\kappa=n-2j$), and of
$M=K^{\cutito}_x+K^{\cutito}_y$ with angular momentum $m$,
placed in the pattern~\eqref{circpat4}. We build these eigenstates
$\ketsub{n,m}{MA}{\circ}$ asking for
\begin{gather}
	K\ketsub{n,m}{MA}{\circ}=(n-2j)\ketsub{n,m}{MA}{\circ},
	\qquad
	M\ketsub{n,m}{MA}{\circ}=m\ketsub{n,m}{MA}{\circ}.
				\label{MM-eigenbasis}
\end{gather}
We observe that whereas the {\sc ra} states in~\eqref{RM-eigenbasis}
are classif\/ied by the Gel'fand--Tsetlin chain of subalgebras
$\Lie{so($4$)}\supset \Lie{so($3$)} \supset \Lie{so($2$)}$, the {\sc
ma} states in \eqref{MM-eigenbasis} follow the chain $\Lie{so($4$)} =
\Lie{su($2$)} \oplus \Lie{su($2$)} \supset \Lie{u($1$)} \oplus
\Lie{u($1$)}$, which in ordinary quantum theory entails the coupling
of two spin-$j$ representations to total spin $\rho$. The overlaps
between the {\sc ma} and {\sc ra} states should therefore be {\it
Clebsch--Gordan coefficients} $\sim C\phantom{|}^{j\,,}_{m_1,}
\phantom{|}^{j\,,}_{m_2,} \phantom{|}^j_m$, with
$m_1=\onehalf(\kappa+m)$ and $m_2=-\onehalf(\kappa-m)$, adding to
the total~$m$~\cite{Bied-Louck}.  In the present construction though,
the subalgebra chain~\eqref{subalg-chain} reduces along the `lower'
subalgebras, and this dif\/fers from the original Gel'fand--Tsetlin
reduction that reduces along the upper ones; also, generally one
counts \Lie{su($2$)} multiplet states from the top down, and we have
ordered them from the bottom up. The overlap of {\sc ra} and {\sc ma}
states entails an extra phase that must be computed carefully. It is
\cite{APVW-II,VW08}
\begin{gather*}
	\!\!\!\!\braketsub{\rho,m}{RA}{{}\ \ \,\circ}{\kappa+j,m}{MA}{\circ}
		 = \varphi(j,\rho,\kappa,m)\,
			C\phantom{|}^{j\,,}_{(m+\kappa)/2,}
			\phantom{|}^{j\,,}_{(m-\kappa)/2,}
					\phantom{|}^{\rho}_{m\mathstrut},
\\
	\varphi(j,\rho,\kappa,m)  := (-1)^{j+\rho}
			\exp[\ii\onehalf\pi	(\kappa+|m|-m)].  
\end{gather*}

\begin{figure}[t]
\centering 
\includegraphics[width=2.2in]{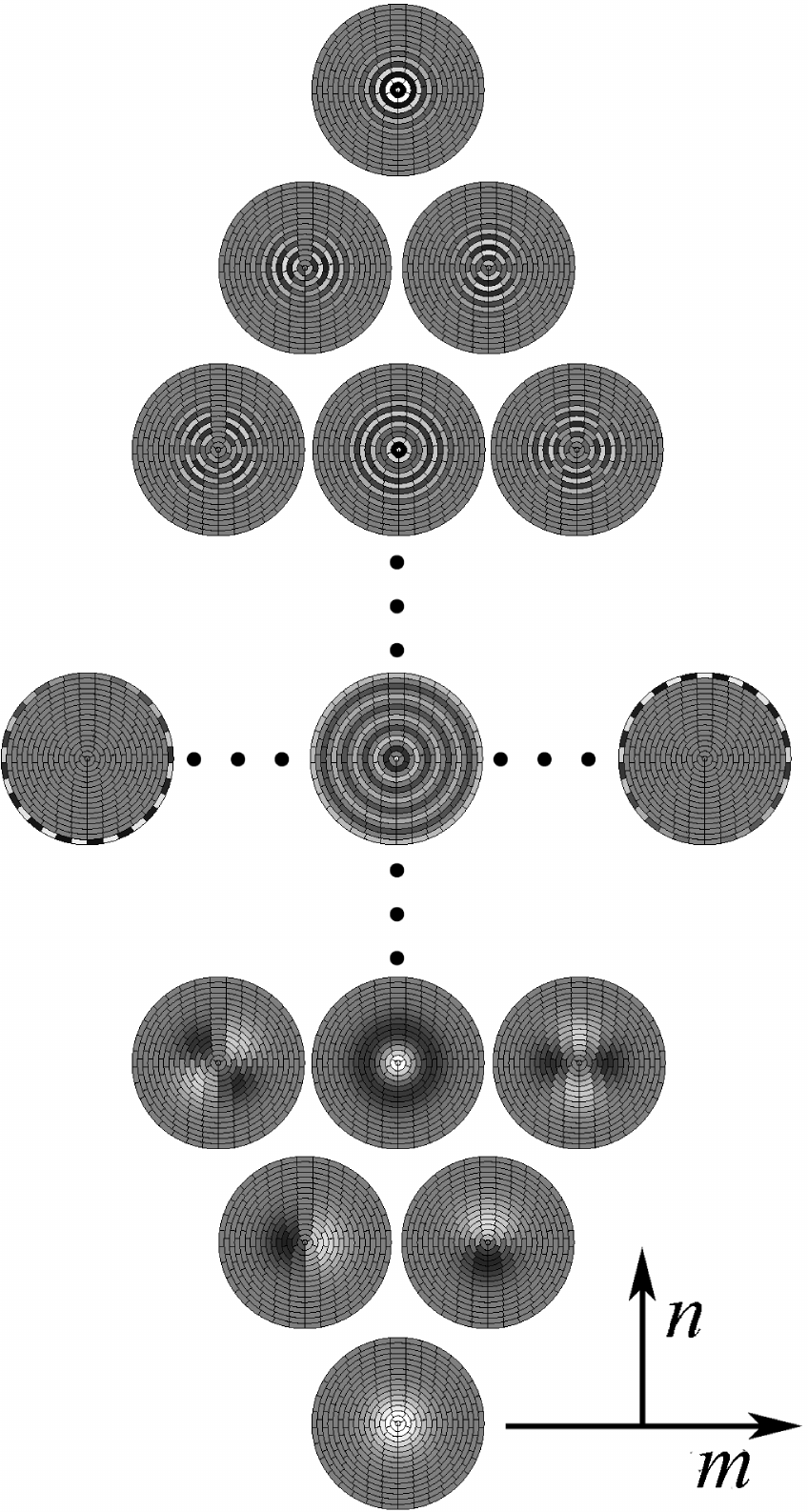}
\caption{Basis of mode-angular momentum eigenstates on the polar
array of radius and angle,
$\Psi^\circ_{n,m}(\rho,\phi_k)$ in~\eqref{r-a-n-m-wavef}.}
\label{fig:base-polar}
\end{figure}

The overlap between the states of mode $n=\kappa+j$ and angular
momentum $m$ with the polar Kronecker basis of radius $\rho$ and
angle $\phi_k$ yields the discrete polar oscillator wavefunctions
	\begin{gather}
	\Psi^\circ_{n,m}(\rho,\phi_k)  :=
		\!\!\!\!\braketsub{\rho,\phi_k}{\odot}{}{n-j,m}{MA}{\circ} \nonumber \\
 \phantom{\Psi^\circ_{n,m}(\rho,\phi_k) }{}\, =
		\frac1{\sqrt{2\rho+1}} \sum_{m=-\rho}^\rho e^{\ii m\phi_k}
			\varphi(j,\rho,\kappa,m)\,
		C\phantom{|}^{j\,,}_{(m+n-j)/2,}\phantom{|}^{j\,,}_{(m-n+j)/2,}
							\phantom{|}^{\rho}_{m\mathstrut} .
  \label{r-a-n-m-wavef}
\end{gather}
These states are accommodated in a rhombus $(n,m)$, similar but {\it
distinct\/} from the rhombus in Fig.~\ref{fig:cuadrado-y-rombo},
which was classif\/ied by $(n_x,n_y)$; here it consists of the eigenvalue
pairs
\begin{gather}
 	n|_0^{2j}, \qquad  m\in\{-n,-n+2,\ldots, n\},\nonumber\\
	n|_{2j}^{4j},\qquad  m\in\{-4j+n,-4j+n+2,\ldots, 4j-n\}.
				\label{nmuranges}
\end{gather}
In Fig.~\ref{fig:base-polar} we show the ${\cal C}^{N^2}$ basis of
{\sc ma} states. Note that due to the Clebsch--Gordan selection rules,
states of a given angular momentum $m$ are nonzero only at radii
$\rho\ge|m|$.

\begin{figure}[t]
\centering 
\includegraphics[width=2.5in]{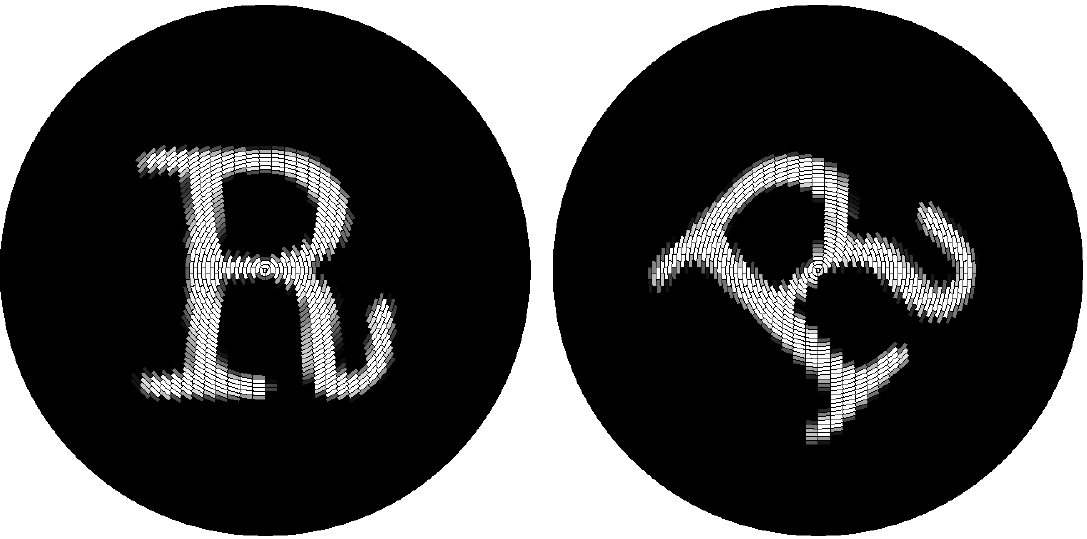}
\caption{Rotation of images on the polar screen.}
\label{fig:polar-rota}
\end{figure}

The generator $K$ in \eqref{circpat4} generates rotations
between both position and momentum operators, and corresponds to
twice the isotropic Fourier transform generator $2\bar F_0$ in
\eqref{symmetricFT}. The action of ${\cal K}(\omega):= \exp(-\ii\omega
K)$ on the {\sc ma} basis of ${\cal C}^{N^2}$ is thus
\begin{gather}
	{\cal K}(\omega)\,\ketsub{n,m}{MA}{\circ}
		=e^{-2\ii(n-2j)\omega}\,\ketsub{n,m}{MA}{\circ}.
					\label{Fou-in-MM}
\end{gather}
Similarly, rotations are generated by angular momentum, ${\cal
R}(\theta):=\exp(-\ii\theta M)$; and since $M$ is twice the generator
$\bar F_3\in\Lie{SU($2$)}_\ssr{F}$ in~\eqref{rotation},
the vectors in the polar basis~\eqref{r-a-n-m-wavef} of ${\cal
C}^{N^2}$ are multiplied by a phase with the double angle,
\begin{gather}
	{\cal R}(\theta)\,\ketsub{n,m}{MA}{\circ}
		=e^{-2\ii m\theta}\,\ketsub{n,m}{MA}{\circ}.
					\label{rot-in-MM}
\end{gather}
Under these rotations, images $f(\rho,\phi_k)=
\brasub{\rho,\phi_k}{\odot}{}\,f\,\rangle$ on the polar-pixellated
screen will thus  transform into
\begin{gather*}
	f_\theta(\rho,\phi_k)  :=  f(\rho,\phi_k+\theta) =
		\brasub{\rho,\phi_k}{\odot}{}{\cal R}(\theta)\,|\,f\rangle
\\
\phantom{f_\theta(\rho,\phi_k)}{} \,	 = \brasub{\rho,\phi_k}{\odot}{}\exp(-\ii\theta M)
	\ketsub{\rho,m}{RA}{\circ}\brasub{\rho,m}{RA}{{}\ \circ}f\rangle
\\
\phantom{f_\theta(\rho,\phi_k)}{}\,	 = \sum_{k'=-\rho}^\rho R^\circ(\rho;\phi_k,\phi_{k'};\theta)
			f(\rho,\phi_{k'})= f(\rho,\phi_k-\theta),
\end{gather*}
where for each radius $\rho|_0^{2j}$ there is a
$(2\rho+1)\times(2\rho+1)$ matrix representing the same rotation
\begin{gather}
	 R^\circ(\rho;\phi_k,\phi_{k'};\theta) :=
	\!\!\!\! \braketsub{\rho,\phi_k}{\odot}{}{\rho,m}{RA}{\circ}
		\exp(-\ii\theta m)\!\!\!
	 \braketsub{\rho,m}{RA}{{}\ \circ}{\rho,\phi_{k'}}{\odot}{}
				\nonumber 
\\
\phantom{R^\circ(\rho;\phi_k,\phi_{k'};\theta)}{}\, 	 =  \frac1{2\rho+1}\sum_{m=-\rho}^\rho
		\exp[-\ii m(\theta-\phi_k+\phi_{k'})] \nonumber\\
\phantom{R^\circ(\rho;\phi_k,\phi_{k'};\theta)}{}\,  = \frac1{2\rho+1}
		\frac{\sin[(\rho+\onehalf)(\theta-\phi_k+\phi_{k'})]
			}{\sin\onehalf(\theta-\phi_k+\phi_{k'})}.
				\label{Rcirc2}
\end{gather}
These are circulating matrices, functions of $\phi_k-\phi_{k'}=
2\pi(k-k')/(2\rho+1)$ modulo $2\pi$, and periodic in $k$, $k'$
modulo $2\rho+1$. For each radius, the `unit' rotation angle is
$\theta=2\pi/(2\rho+1)$, and for multiples $l$ thereof, the matrix
\eqref{Rcirc2} is nonzero at the diagonal $k=k'+l$. In Fig.~\ref{fig:polar-rota} we give an example of such rotation.

Isotropic Fourier transformations and rotations on the polar screen
are produced by generators within the \Lie{so($4$)} algebra in the
pattern~\eqref{circpat4}. However, the pattern also informs us that with
linear combinations of $K$ and $M$, we can {\it not\/} gyrate the
planes $(Q^\circ_x,P^\circ_y)$ and $(Q^\circ_y,P^\circ_x)$ jointly as
with $2\bar F_2$ in \eqref{gyration}; also missing is the anisotropic Fourier
transform generated by $2\bar F_1$ in~\eqref{skewsymmetricFT}, which was
natural in the Cartesian basis. These transformations will be {\it
imported\/} on~${\cal C}^{N^2}$ in Section \ref{sec:seven}.

\section{Importation of rotations on the Cartesian screen}
											\label{sec:five}

	 We noted above that in the subalgebra chain \eqref{subalg-chain},
the generators of isotropic Fourier transformations and rotations,
$K\leftrightarrow2\bar F_0$ and $M\leftrightarrow2\bar F_3$ are {\it
domestic\/} to \Lie{so($4$)}, while anisotropic Fourier
transformations and gyrations, corresponding to $2\bar F_2$, are {\it
foreign}.  Now, in the Cartesian subalgebra decomposition of
\Lie{so($4$)} in \eqref{direct-sum}, the two independent Fourier
transform generators \eqref{symmetricFT},~\eqref{skewsymmetricFT},
$K^{\cutito}_x\leftrightarrow \bar F_0+\bar F_1$ and
$K^{\cutito}_y\leftrightarrow \bar F_0-\bar F_1$ are {\it domestic\/} to
\Lie{so($4$)}, while those of gyrations and rotations, $2\bar F_2$ and
$2\bar F_3$, are {\it foreign}. Since we cannot complete
a fully domestic $\Lie{U($2$)}_\ssr{K}$ Fourier--Kravchuk group in
correspondence with the Fourier group $\Lie{U($2$)}_\ssr{F} \subset
\Lie{Sp($4,\Re$)}$, we must {\it import\/} the missing
transformations. Such importation was used in \eqref{rho-phi-basis} with
the $(2\rho+1)\times (2\rho+1)$ discrete Fourier transform matrix.

We now build the group of rotations on the Cartesian
grid by importing \Lie{SU($2$)} transformations \cite[Chapter~3]{Bied-Louck} from the continuous model.  Rotations of an image
should respect the energy of each formant mode $n=n_x+n_y=\kappa+2j$
in the $\ketsub{n_x,n_y}{3}{\cutito}$ basis of ${\cal C}^{N^2}$, and
transforms real images into real ones, while mixing states with
dif\/ferent eigenvalues $\mu:=\onehalf(n_x-n_y)$, $\mu|_{-n/2}^{n/2}$,
horizontally across the rhombus in Fig.~\ref{fig:cuadrado-y-rombo}{\it b}. The proposed imported action
of $\bar F_3$ in \eqref{rotation} on the Cartesian modes stems from~\eqref{matL}, replacing $j\mapsto \onehalf(n_x+n_y)= \onehalf n$
and $m \equiv \mu \mapsto \onehalf(n_x-n_y)$, namely
\begin{gather}
	M \ketsub{n_x,n_y}{3}{\cutito}
	= \sqrt{n_y(n_x+1)}\ketsub{n_x+1,n_y-1}{3}{\cutito}
	+ \sqrt{n_x(n_y+1)}\ketsub{n_x-1,n_y+1}{3}{\cutito}.
					\label{actFrot}
\end{gather}
We must pay attention to the fact that $\bar F_3$ in \eqref{rotation} is {\it
one-half\/} of the angular momentum operator $q_xp_y-q_yp_x$ that is
the generator of f\/inite rotations, ${\cal R}(\theta) \leftrightarrow
\exp(-2\ii\theta \bar F_3)$. On the Cartesian mode basis states its action
involves the standard Wigner little-$d$ function \eqref{func-d} through
\begin{gather*}
	{\cal R}(\theta)\ketsub{n_x,n_y}{3}{\cutito}
	=\sum_{n'_x+n'_y=n} d^{n/2}_{(n_x-n_y)/2,(n'_x-n'_y)/2}(2\theta)
		\ketsub{n'_x,n'_y}{3}{\cutito}.
\end{gather*}
This is a real linear combination of the Cartesian mode basis
where $n$ and $\mu$ are bound within the rhombus of Fig.~\ref{fig:cuadrado-y-rombo}{\it b}, for integer
$n|_0^{2j}$, $\mu|_{-n/2}^{n/2}$ in the lower half and for
$n|_{2j}^{4j}$, $\mu|_{n/2-j}^{2j-n/2}$ in the upper one.

\begin{figure}[t]
\centering 
\includegraphics[width=2.5in]{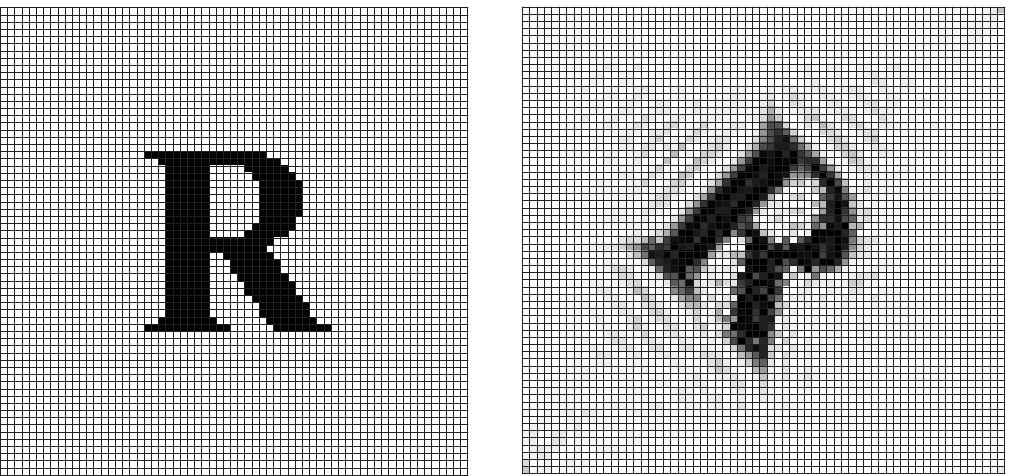}
\caption{Rotation of images on the square screen.}
\label{fig:Carte-rota}
\end{figure}

In Fig.~\ref{fig:Carte-rota} we show rotation of a
Cartesian-pixellated image $f(q_x,q_y)=
\brasub{q_x,q_y}{1}{}f\rangle$ to
\begin{gather*}
	f_\theta(q_x,q_y) :=
		\brasub{q_x,q_y}{1}{\cutito}{\cal R}(\theta)\,|f\rangle
  = \brasub{q_x,q_y}{1}{\cutito}{\cal R}(\theta)\,
				\ketsub{n_x,n_y}{3}{\cutito}
			\braketsub{n_x,n_y}{3}{\cutito}{q'_x,q'_y}{1}{\cutito}
						\brasub{q'_x,q'_y}{1}{\cutito}f\rangle
						\nonumber\\
\phantom{f_\theta(q_x,q_y)}{}\, =  \sum_{q'_x,q'_y} R^{\cutito}(q_x,q_y;q'_x,q'_y;\theta)
							f(q'_x,q'_y).
	\end{gather*}
The subgroup of rotations ${\cal R}(\theta)\in\Lie{SU($2$)}_\ssr{K}$
is thus represented by the $N^2\times N^2$ matrices
\begin{gather}
	R^\cutito(q_x,q_y;q'_x,q'_y;\theta)  := \!\!\!\braketsub{q_x,q_y}{1}{\cutito}{n_x,n_y}{3}{\cutito}
     \brasub{n_x,n_y}{3}{\cutito}{\cal R}(\theta)
						\ketsub{n'_x,n'_y}{3}{\cutito}
			\!\!\braketsub{n'_x,n'_y}{3}{\cutito}{q'_x,q'_y}{1}{\cutito}
						\nonumber 
\\
\phantom{R^\cutito(q_x,q_y;q'_x,q'_y;\theta) }{}\, =\sum_{\mu,\mu'}
		\Psi^\cutito_{n_x,n_y}(q_x,q_y)
				 d^{n/2}_{\mu,\mu'}(2\theta)
		\Psi^\cutito_{n'_x,n'_y}(q'_x,q'_y),
								\label{RF5}
\end{gather}
where $\mu=\onehalf(n_x-n_y)$ and $\mu'=\onehalf(n'_x-n'_y)$ are
bound by $n_x+n_y=n=n'_x+n'_y$, and belong to the same row in the
rhombus of Fig.~\ref{fig:cuadrado-y-rombo}{\it b}.

\section[Completion of $\Lie{U($2$)}_\ssr{K}$ on the Cartesian screen]{Completion of $\boldsymbol{\Lie{U($2$)}_\ssr{K}}$ on the Cartesian screen}

								\label{sec:six}

Having imported a unitary representation of the group of rotations
onto pixellated images on the Cartesian screen, and having the
domestic group of fractional Fourier transforms, we can complete the
Fourier--Kravchuk group $\Lie{U($2$)}_\ssr{K}$ on this screen.
From \eqref{skewsymmetricFT} and \eqref{direct-sum},
$\bar F_1\leftrightarrow\onehalf(K^\cutito_x-K^\cutito_y)$, the Lie
exponential ${\cal A}(\phi)\leftrightarrow\exp(-2\ii\phi \bar F_1)$
acts on the Cartesian kets \eqref{base-kxky} through phases,
\begin{gather}
	{\cal A}(\phi) \ketsub{n_x,n_y}{1}{\cutito}
			= \exp[-2\ii\phi(n_x-n_y)]
			\ketsub{n_x,n_y}{1}{\cutito}.
					\label{skewsymmact}
\end{gather}
Hence, for images $f(q_x,q_y)$,
	\begin{gather*}
	f_\phi(q_x,q_y) :=  {\cal A}(\phi)\,f(q_x,q_y)
			 = 	 \sum_{q'_x,q'_y} A^\cutito(q_x,q_y;q'_x,q'_y;\phi)
							f(q'_x,q'_y),
\end{gather*}
with the matrix kernel
\begin{gather}
	A^\cutito(q_x,q_y;q'_x,q'_y;\phi)=\sum_{\mu}
		\Psi^\cutito_{n_x,n_y}(q_x,q_y)
				 \exp[-2\ii\phi(n_x-n_y)]
		\Psi^\cutito_{n_x,n_y}(q'_x,q'_y),
								\label{RFA}
\end{gather}
where the sum over $\mu=\onehalf(n_x-n_y)$ preserves~$n$.

Now, having two of the three generators of $\Lie{SU($2$)}_\ssr{K}$,
we can produce the third: gyrations ${\cal G}(\psi)\leftrightarrow
\exp(-2\ii\psi \bar F_2)$, through
\begin{gather*}
	{\cal G}(\psi)={\cal A}(\tsty14\pi) {\cal R}(\psi)
					{\cal A}(\tsty14\pi)^{-1}.
\end{gather*}
On images $f(q_x,q_y)$, Fourier--Kravchuk gyrations will act through a
matrix kernel,
\begin{gather*}
	f_\psi(q_x,q_y)  :=  {\cal G}(\psi)\,f(q_x,q_y)
			=\sum_{q'_x,q'_y} G^\cutito(q_x,q_y;q'_x,q'_y;\psi)
							f(q'_x,q'_y),		
\\
	G^\cutito(q_x,q_y;q'_x,q'_y;\psi)
	 = \sum_{\mu,\mu'}
		\Psi^\cutito_{n_x,n_y}(q_x,q_y)
			e^{-\ii\pi \mu/4}\,
			d^{n/2}_{\mu,\mu'}(2\psi)
			e^{+\ii\pi \mu'/4}
		\Psi^\cutito_{n'_x,n'_y}(q'_x,q'_y),
\end{gather*}
where, as in \eqref{RF5} and \eqref{RFA}, the sums over
$\mu=\onehalf(n_x-n_y)$ and $\mu'=\onehalf(n'_x-n'_y)$ preserve~$n$.
In continuum optics, gyrations acting on
the Hermite--Gauss beams transform them into La\-guerre--Gauss ones of
the same mode number \cite[Fig.~4]{Madrid}.

Finally, the isotropic ${\cal K}(\omega)\leftrightarrow
\exp(-2\ii\omega \bar F_0)$ in~\eqref{Fou-in-MM} is domestic to the \Lie{so($4$)} algebra~\eqref{direct-sum}, and completes the $\Lie{U($2$)}_\ssr{K}$ group with
\begin{gather*}
	f_\omega(q_x,q_y) :=  {\cal K}(\omega)\,f(q_x,q_y)
	=\sum_{q'_x,q'_y} K^\cutito(q_x,q_y;q'_x,q'_y;\omega) 	f(q'_x,q'_y),
\\
	K^\cutito(q_x,q_y;q'_x,q'_y;\omega)
		 = \sum_{n_x,n_y}
		\Psi^\cutito_{n_x,n_y}(q_x,q_y)
				\,\exp[-2\ii\omega(n_x+n_y)]
		\Psi^\cutito_{n_x,n_y}(q'_x,q'_y).
\end{gather*}

The elements of the Fourier group $\Lie{U($2$)}_\ssr{F} =
\Lie{U($1$)}_\ssr{F} \otimes \Lie{SU($2$)}_\ssr{F}$, where the
factors are complementary, are customarily parametrized by Euler
angles as
	\begin{gather}
	\bar{\cal D}(\omega;\phi,\theta,\psi)  :=   \exp(-\ii\omega \bar F_0)
  \exp(-\ii\phi \bar F_3)
		 \exp(-\ii\theta \bar F_2) \exp(-\ii\psi \bar F_3),
						\label{gen-calD}
\end{gather}
and its matrix elements between eigenstates $\brasub{\iota,\mu}{}{}$
and $\ketsub{\iota,\mu'}{}{}$, with eigenvalues $\mu$, $\mu'$ under
$\bar F_3$ and $\iota$ under $\bar F_0$, the latter being the irreducible
representation label are the well-known Wigner {\it Big-D} functions
\begin{gather*}
	D^\iota_{\mu,\mu'}(\omega;\phi,\theta,\psi)
		=e^{-\ii\iota\omega} e^{-\ii\mu\phi}
				d^\iota_{\mu,\mu'}(\theta) e^{-\ii\mu'\psi}.
\end{gather*}
As we indicated in Section~\ref{sec:three}, by permuting $1\mapsto2
\mapsto3 \mapsto1$ in \eqref{gen-calD} and using~\eqref{Fou-in-MM},
\eqref{rot-in-MM}, and~\eqref{skewsymmact}, we can write the elements of
the isomorphic Fourier--Kravchuk group $\Lie{U($2$)}_\ssr{K}$ as a~pro\-duct of Fourier--Kravchuk transforms and rotations,
\begin{gather*}
	{\cal D}(\omega;\phi,\theta,\psi)=
		{\cal K}(\onehalf\omega)  {\cal A}(\onehalf\phi)
				{\cal R}(\onehalf\theta)
				{\cal A}(\onehalf\psi).
\end{gather*}
Its matrix elements between the Cartesian mode eigenstates
$\ketsub{n_x,n_y}{1}{\cutito}$ of $K^{\cutito}_x$ and $K^{\cutito}_y$will be thus,
\begin{gather}
	\brasub{n_x,n_y}{1}{\cutito}{\cal D}(\omega;\phi,\theta,\psi)
				\ketsub{n'_x,n'_y}{1}{\cutito}
	= e^{-\ii(n-2j)\omega}
		D^{n/2}_{(n_x-n_y)/2,(n'_x-n'_y)/2}(\phi,\theta,\psi),
				\label{Duaga}
\end{gather}
with the total mode $n=n_x+n_y=n'_x+n'_y$ as before.

The domestic and imported transformations properly mesh, and we see
that the Fourier--Kravchuk group $\Lie{U($2$)}_\ssr{K}$ is indeed
represented unitarily and faithfully by \eqref{Duaga} on ${\cal
C}^{N^2}$. Its action on
images $f(q_x,q_y)$ pixellated on the Cartesian screen can be found
from here, writing $\Omega:=(\phi,\theta,\psi)$, through a real
similarity transformation by the matrix formed with the
${\cal C}^{N^2}$ basis~\eqref{Twodimpsi} of Cartesian mode Kravchuk
functions,
\begin{gather}
	f_{\omega,\Omega}(q_x,q_y):= {\cal D}(\omega,\Omega)\,f(q_x,q_y)
			 =\sum_{q'_x,q'_y}
		D^\cutito(q_x,q_y;q'_x,q'_y;\omega,\Omega)\,f(q'_x,q'_y),
						\label{OOmegaq}
\end{gather}
where, with $n=n_x+n_y=n'_x+n'_y$ and $n|_0^{4j}$ the kernel is
  \begin{gather}
	D^\cutito(q_x,q_y;q'_x,q'_y;\omega,\Omega)\nonumber\\
\qquad {} =\sum_{\textstyle{n_x,n_y\atop n'_xn'_y}}
		\Psi^\cutito_{n_x,n_y}(q_x,q_y)\,e^{-\ii(n-2j)\omega}
		\,D^{n/2}_{(n_x-n_y)/2,(n'_x-n'_y)/2}(\Omega)
		\Psi^\cutito_{n'_x,n'_y}(q'_x,q'_y),
						\label{DOOmegaq}
\end{gather}
and these $N^2\times N^2$ matrices are unitary representations of
\Lie{U($2$)} on the Cartesian grid of points in Fig.~\ref{fig:cuadrado-y-rombo}{\it b}.
With this result we now turn to the polar screen.

\section{The Fourier group on polar screens}
								\label{sec:seven}

In the subalgebra chain of \Lie{so($4$)} that produces the polar
screen \eqref{circpat4}, the commuting rotations and isotropic
Fourier--Kravchuk transforms are domestic. To complete the
Fourier--Kravchuk group $\Lie{U($2$)}_\ssr{K}$ on the polar screen, we
must import either gyrations or the anisotropic transform. Since
these transformations have been realized already in the Cartesian
screen basis, we should f\/ind a unitary map between both screens. This
transformation has been studied in \cite{APVW-II,VW08},
and consists in identifying the eigenstates of mode and angular
momentum in both bases; in the polar basis these are
$\ketsub{\rho,m}{RA}{\circ}$ in \eqref{RM-eigenbasis} shown in Fig.~\ref{fig:base-polar}. In the Cartesian basis such states will be
constructed now as eigenvectors of ${\cal R}(\theta)=\exp(-\ii\theta
M)$ with eigenvalues $e^{-\ii\theta m}$, or equivalently of $\bar
F_3$ with eigenvalues $\onehalf m$, corresponding to the eigenvalues
of $\onehalf M$ in \eqref{rotation}. These Cartesian states will be
linear combinations, respecting total mode number $n=n_x+n_y$, of all
states $\ketsub{n_x,n_y}{1}{\cutito}$,
\begin{gather*}
 \ketsub{n,m}{MA}{\cutito}:=
		\sum_{n_x+n_y=n} C^{n,m}_{n_x,n_y}
		\ketsub{n_x,n_y}{1}{\cutito}, 
\\
  M \ketsub{n,m}{MA}{\cutito}
		=  m \ketsub{n,m}{MA}{\cutito},
			\qquad K\ketsub{n,m}{MA}{\cutito}
			=\onehalf n \ketsub{n,m}{MA}{\cutito}.
\end{gather*}
From \eqref{actFrot}, the coef\/f\/icients obey the dif\/ference equation
\begin{gather*}
	\sqrt{n_y(n_x+1)} C^{n,m}_{n_x+1,n_y-1}
	-\onehalf m\,C^{n,m}_{n_x,n_y}
	+\sqrt{n_x(n_y+1)} C^{n,m}_{n_x-1,n_y+1} =0,
\end{gather*}
which is the ubiquitous \Lie{su($2$)} three-term recursion relation
of the Wigner little-$d$ functions for angle
$\onehalf\pi$ (around the 1-axis \cite{Bied-Louck}), now with
$j\leftrightarrow\onehalf n=\onehalf(n_x+n_y)$ and $\onehalf
m\leftrightarrow\onehalf(n_x-n_y)$. We thus def\/ine {\sc ma} states
having angular momentum $m$ on the square grid as
\begin{gather}
	\ketsub{n,m}{MA}{\cutito}:=\sum_{n_x+n_y=n}
	e^{\ii\pi(n_x-n_y)/4} d^{n/2}_{m/2,(n_x-n_y)/2}(\onehalf\pi)
			\ketsub{n_x,n_y}{1}{\cutito}. \label{finketnmu}
\end{gather}
These states can be accomodated in a rhombus $(n,m)$, exactly as
in~\eqref{nmuranges}.

\begin{figure}[t]
\centering 
\includegraphics[width=2.2in]{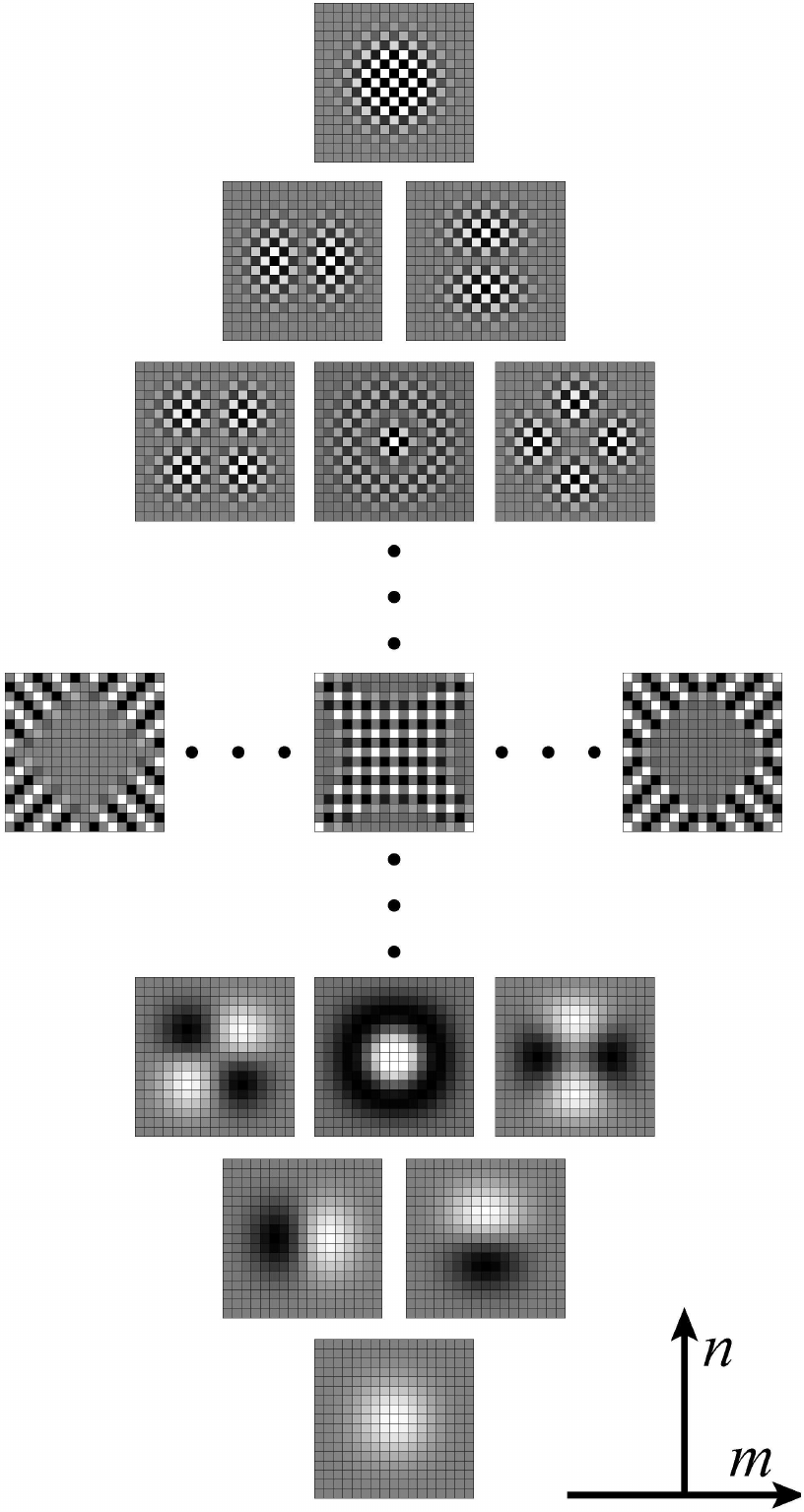}
\caption{Basis of mode-angular momentum eigenstates on the Cartesian
array, $\Lambda_{n,m}^{\cutito}(q_x,q_y)$ in \eqref{finketnmu}.}
\label{fig:E-AM-Cart}
\end{figure}

From \eqref{finketnmu} follows the def\/inition of the Cartesian basis of
mode and angular momentum {\sc ma} eigenstates
\begin{gather}
	\Lambda_{n,m}^{\cutito}(q_x,q_y) :=
			\braketsub{q_x,q_y}{1}{\cutito}{n,m}{MA}{\cutito}
	 						\label{defLambda}\\
\phantom{\Lambda_{n,m}^{\cutito}(q_x,q_y)}{} \, = \sum_{n_x+n_y=n}
	e^{\ii\pi(n_x-n_y)/4} d^{n/2}_{m/2,(n_x-n_y)/2}(\onehalf\pi)
			\Psi_{n_x,n_y}^{\cutito}(q_x,q_y) \nonumber\\ 
\phantom{\Lambda_{n,m}^{\cutito}(q_x,q_y)}{} \, = \Lambda_{n,-m}^{\cutito}(q_x,q_y)^*
				  = (-1)^{n+m}\Lambda_{n,m}^{\cutito}(-q_x,-q_y)
=  (-1)^{q_x+q_y}\Lambda_{4j-n,-m}^{\cutito}(q_x,q_y).
				  			\nonumber 
\end{gather}
This transforms the Cartesian Kravchuk functions into what we called
La\-guerre--Kravchuk functions in~\cite[Fig.~4]{Madrid}.
The basis of states \eqref{defLambda} is shown in Fig.~\ref{fig:E-AM-Cart}, which
can now be identif\/ied with the basis~-- also of mode and angular
momentum~-- in Fig.~\ref{fig:base-polar}.
We can now {\it import\/} the equivalence between the {\sc ma} bases
\begin{gather}
	\ketsub{n,m}{MA}{\cutito} \equiv \ketsub{n,m}{MA}{\circ}.
				\label{C-p-equiv}
\end{gather}

\begin{figure}[t]
\centering
\includegraphics[width=2.5in]{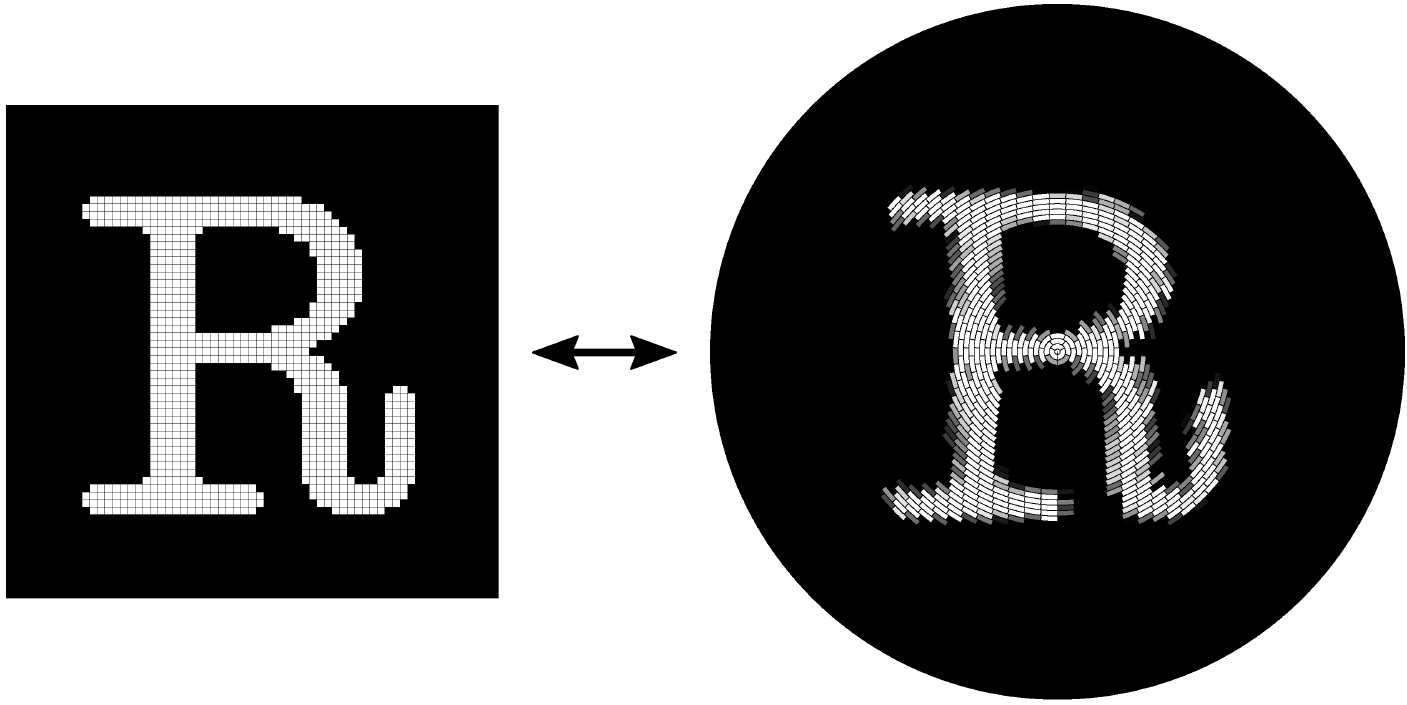}
\caption{Unitary map~\eqref{fUcuadr} of an image (the letter R) on the
Cartesian screen onto the polar screen.}
\label{fig:R-cuad-circ}
\end{figure}

A pixellated image on the  Cartesian screen, $f_{\cutito}(q_x,q_y) \in
{\cal C}^{N^2}$, can be thus unitarily transformed into an image on
the polar screen, $f_\circ(\rho,\phi_k) \in {\cal C}^{N^2}$ through
the transformation
\begin{gather}
	f_\circ(\rho,\phi_k) =  \brasub{\rho,\phi_k}{\odot}{}f\rangle
									\nonumber\\ 
\phantom{f_\circ(\rho,\phi_k)}{} 	 = \sum_{q_x,q_y}\!\!\!
		\braketsub{\rho,\phi_k}{\odot}{}{q_x,q_y}{1}{\cutito}
		\brasub{q_x,q_y}{1}{\cutito}f\rangle   =  \sum_{q_x,q_y} U(\rho,\phi_k;q_x,q_y)
				f_{\cutito}(q_x,q_y),
							\label{fUcuadr}
\end{gather}
where the transform kernel, using~\eqref{C-p-equiv}, is
\begin{gather*}
	U(\rho,\phi_k;q_x,q_y) :=\!\!\!\!
		\braketsub{\rho,\phi_k}{\odot}{}{q_x,q_y}{1}{\cutito}
									\\ 
\phantom{U(\rho,\phi_k;q_x,q_y)}{}\, 	 \equiv \sum_{n,m}\!\!\!
		\braketsub{\rho,\phi_k}{\odot}{}{n,m}{MA}{\circ}
		\braketsub{n,m}{MA}{\cutito}{q_x,q_y}{1}{\cutito}
									  = \sum_{n,m}
		\Psi^\circ_{n,m}(\rho,\phi_k)
				\Lambda_{n,m}^{\cutito}(q_x,q_y)^*.
\end{gather*}
The transformation of images inverse to~\eqref{fUcuadr}, from the polar
to the Cartesian screen, is
\begin{gather*}
	f_{\cutito}(q_x,q_y) =
	\sum_{\rho,\phi_k} V(q_x,q_y;\rho,\phi_k)
				f_\circ(\rho,\phi_k),
									\\ 
	V(q_x,q_y;\rho,\phi_k)
			 = \sum_{n,m} \Lambda_{n,m}^{\cutito}(q_x,q_y)
				\Psi^\circ_{n,m}(\rho,\phi_k)^*
				=U(\rho,\phi_k;q_x,q_y)^*.
\end{gather*}
We note that these transformations map real functions onto real
functions. In Fig.~\ref{fig:R-cuad-circ} we show an example of the
map \eqref{fUcuadr} on the 0-and-1 image of the letter~R.

The action of the Fourier--Kravchuk group $\Lie{U($2$)}_\ssr{K}$ on
images on the polar screen can be now derived from its action
\eqref{OOmegaq}, \eqref{DOOmegaq} on the Cartesian screen:
\begin{gather*}
	f_{\omega,\Omega}(\rho,\phi_k):=
		{\cal D}(\omega,\Omega)\,f(\rho,\phi_k)
			 =\sum_{\rho',\phi_{k'}}
		D^\odot(\rho,\phi_k;\rho',\phi_{k'};\omega,\Omega)
					f(\rho',\phi_{k'}),
\end{gather*}
where the kernel is a representation of $\Lie{U($2$)}_\ssr{K}$ on the
polar screen
\begin{gather*}
	D^\odot(\rho,\phi_k;\rho',\phi_{k'};\omega,\Omega)
		 =\sum_{\textstyle{q_x,q_y\atop q'_x,q'_y}}
			U(\rho,\phi_k;q_x,q_y)
			D^\cutito(q_x,q_y;q'_x,q'_y;\omega,\Omega)
			V(\rho',\phi_{k'};q'_x,q'_y).
\end{gather*}
We have thus far not found a more compact analytic form for the
representation of the Fourier group on the discrete coordinates of
radius and angle.

\looseness=-1
The $N^2\times N^2$ matrix $U =
(V)^\dagger$ that is needed above, is arduous to
calculate but its numerical values need be computed only once for
each size $N$ of the screen, and will serve to transform any image
from the Cartesian to the polar screen and back. For $N=32$ there are
$\approx10^6$ values to be stored, and the transformation of an image
as that in Fig.~\ref{fig:R-cuad-circ} between the screens involves
this number of sums and products. Commercial image-rotating
algorithms store the original image and interpolate the few Cartesian
pixels around each geometrically determined point. Certainly, the
Fourier--Kravchuk group of transformations of images presented here
does not provide a fast algorithm, but it is unitary, and hence does
not loose information, as interpolation invariably does; it is exact.

\section{Concluding remarks}\label{sec:eight}

Willard Miller Jr.\ is recognized as having initiated the modern
study of systems whose governing equation allows for separation in
various coordinate systems using Lie algebraic reduction methods
\cite{Miller-book}. These equations are dif\/ferential because space
has been considered continuous.  Finite spaces with discrete
coordinates, used as foundation for f\/inite Hamiltonian systems
characterized by governing dif\/ference equations, have not been
considered~-- to our knowledge~-- in the same context. The present
paper reviews our work in the particular case of a two-dimensional
Hamiltonian system whose two governing Hamilton equations identify as
a f\/inite harmonic oscillator, mothered by the compact algebra
\Lie{so($4$)}. And since it happens that algebra has two distinct
subalgebra reductions, we can relate them with Cartesian an polar
coordinates.

\looseness=-1
The dif\/ference equations whose solutions are the f\/inite oscillator
wavefunctions and the Clebsch--Gordan coef\/f\/icients are contained in
\Lie{so($3$)} and \Lie{so($4$)}, and entail special Kravchuk and Hahn
polynomials. Other f\/inite polynomials, such as those of Meixner and
Pollaczek, also appear in related harmonic oscillator models
\cite{AW-Meixner-osc}, and thus could be interpreted in terms of
two-dimensional physical and optical models. Further, f\/inite
three-dimensional coordinate systems beyond Cartesian and circular
cylindric ones may be of interest in applications.  But we should end
with a word of caution regarding applications: the two-dimensional
arrays that purportedly measure the quality of Laguerre--Gauss (polar)
laser beams generally use {\it sampled\/} Laguerre--Gauss functions;
the question of whether these, or the mode-angular momentum f\/inite
functions, are the most ef\/f\/icient to f\/ind the mode and angular
momenta of the actual beams, seems to favor the sampled functions
\cite{approx,approx2}, although they do not form orthonormal bases
for the space of pixellated images. But only the discrete function
bases {\it synthesize\/} all images exactly.

The continuous two-dimensional harmonic oscillator is a
superintegrable system, which is known to separate generally in
elliptic coordinates, of which the Cartesian and polar are limits. We
have been unable so far to f\/ind a corresponding f\/inite array that
would allow for pixellation following ellipses and hyperbolas,
although there are three enticing leads: diagonalization of the
\Lie{so(3)} operator $J^2+\alpha J_3$ \cite{Miller-book}; a line of
functions with a parameter which joins continuously Kronecker deltas
with Clebsch--Gordan coef\/f\/icients \cite{Pogosyan}; and the production
by holographic means of laser beams whose nodes follow elliptic
coordinates \cite{Julio-Cesar-GV}. There are also `geometrical'
reasons to doubt that this is possible, however: whereas in Cartesian
coordinates all pixel sizes are equal, and in polar ones they are
almost so (except very near to the center), trying to visualize these
two pixellations as limits of a f\/inite elliptic pixellation presents
some problems that we have not yet been able to overcome \cite{fail}.

\subsection*{Acknowledgements}

We thank the support of the {\it \'Optica Matem\'atica} projects
DGAPA-UNAM IN-105008 and SEP-CONACYT 79899, and we thank Guillermo
Kr\"otzsch (ICF-UNAM) for his assistance with the graphics and Juvenal Rueda-Paz (Facultad de Ciencias, Universidad Aut\'onoma del Estado de Morelos) for his support with the manuscript.

\pdfbookmark[1]{References}{ref}
\LastPageEnding

\end{document}